\documentclass[5ptstructabstract]{aa}  
\usepackage{natbib}
\usepackage{lscape}
\usepackage{rotating}
\usepackage[T1]{fontenc}
\usepackage[utf8]{inputenc} 
\usepackage[toc,page]{appendix}
\usepackage{graphicx}
\usepackage{enumerate}
\usepackage{amsmath}
\usepackage{subcaption}
\usepackage{txfonts}
\usepackage{float}
\usepackage{textcomp}
\usepackage{gensymb}
\usepackage[colorlinks=true,linkcolor=blue,citecolor=blue]{hyperref}
\usepackage{color}
\usepackage{mathtools}
\usepackage{mathrsfs}
\usepackage[para,online,flushleft]{threeparttable}

\newcommand{\mesa}{{\tt MESA}}
\newcommand{\Msun}{\ensuremath{{\rm M}_\odot}}
\newcommand{\devi}[1]{{\color{black}#1}}

\begin{document}

  \title{Investigating cannibalistic millisecond pulsar binaries using \mesa\,: New constraints from pulsar spin and mass evolution
\\}

   \author{Devina Misra
          \inst{1}\fnmsep\thanks{e-mail: devina.misra@ntnu.no},
          Manuel Linares\inst{1,2},
          Claire S. Ye\inst{3}
          }

   \authorrunning{Misra et al.}
   \titlerunning{Investigating cannibalistic millisecond pulsar
   binaries}

   \institute{Department of Physics, Norwegian University of Science and Technology, NO-7491 Trondheim, Norway
              \and
            Departament de F{\'i}sica, EEBE, Universitat Polit{\`e}cnica de Catalunya, Av. Eduard Maristany 16, E-08019 Barcelona, Spain.
              \and
             Canadian Institute for Theoretical Astrophysics, University of Toronto, Toronto, Ontario M5S 3H8, Canada
             }

   \date{Received 28/08/2024; Accepted 14/12/2024}

 
  \abstract
    {Compact binary millisecond pulsars (MSPs) with orbital periods $\lesssim1$d are key to understanding binary evolution involving massive neutron stars (NSs). Due to the ablation of the companion by the rapidly spinning pulsar, these systems are also known as spiders and categorized into two main branches: redbacks (RBs; companion mass in the range of 0.1 to 0.5\,\Msun) and black widows (BWs; companion mass $\lesssim$\,0.1\,\Msun).}
   {We present models of low- and intermediate-mass X-ray binaries and compare them with observations of Galactic spiders (including the presence or absence of hydrogen lines in their optical spectra), and we constrain and quantify the interaction between the pulsar and the companion.
   }
   {Using \mesa, we created the allowed initial parameter space. For the first time in \mesa\,, we also included the detailed evolution of the pulsar spin \devi{and modeled the irradiation of the companion by the pulsar wind}. } 
   {Efficient mass accretion onto the NS (i.e., at least $70\%$ of the mass transferred is accreted) with an X-ray irradiated disk followed by strong irradiation of the companion can explain most of the properties of the observed spiders. Our RB evolutionary tracks continue to the BW regime, connecting the two branches of spiders. Our models explain the lack of hydrogen in some observed BWs with ultra-light companions. During accretion induced spin up, the mass required to spin up an NS to sub-milliseconds is high enough to collapse it into a black hole. Finally, after analyzing the formation of RB-like spiders with giant companions and orbital periods of several days (huntsmen), we conclude that they are unlikely to produce super-massive NSs (maximum accreted mass $\lesssim$0.5 $\Msun$).
   }
   {Cannibalistic MSP binary formation depends heavily on the interplay between accretion onto the pulsar and pulsar wind irradiation.
   Our work supports earlier claims that RBs evolve into BWs.
   We also show that the fastest spinning pulsars may collapse before reaching sub-millisecond spin periods. 
   
   }


   \maketitle

\section{Introduction}

The formation of radio millisecond pulsars (MSPs) has long been connected to low-mass X-ray binaries (LMXBs; \citep{1982Natur.300..728A,8812f611-e297-3575-a70c-49607c391972,1991PhR...203....1B}). Observed MSPs have short spin periods ($\lesssim 30$~ms) and low surface magnetic fields ($\sim 10^{8}$~G). The neutron stars (NSs) in LMXBs accrete matter that spins them up to millisecond periods \citep{1982Natur.300..728A}. This phase is observed as an X-ray phase and is also called the recycling phase due to the pulsar spin-up. Observations of accreting X-ray MSPs confirmed the link between MSPs and accreting LMXBs \citep{1998Natur.394..344W,2009Sci...324.1411A}. The NS surface magnetic field has also been suggested to decrease with the accreted mass due to Ohmic decay, explaining the observed magnetic field strengths of MSPs \citep{1997MNRAS.284..311K,1998A&A...330..195Z,1999MNRAS.303..588K, 1999MNRAS.308..795K, 2001ApJ...557..958C}. Compact binary MSPs are a sub-category of MSP binaries with compact orbits and eclipses in their radio signals. In compact binary MSPs, after the mass-transfer phase or Roche-lobe overflow (RLO) has ended, the highly spinning pulsar loses energy in the form of energetic particles that irradiate the outer surface of the companion, leading to mass loss \citep{2009Sci...324.1411A}. Initial observations of eclipsing MSPs showed that some of them have very low mass companions \citep[0.02 to 0.05\,\Msun;][]{1988Natur.333..237F,1996ApJ...465L.119S,1999ApJ...510L..45S}, which could be explained by mass loss due to pulsar wind irradiation. Evidence of irradiation-induced mass loss has since been observed in multiple eclipsing compact MSPs \citep{1988Natur.334..225K, 1989ApJ...336..507R,1989ApJ...343..292R, 1996ApJ...473L.119S,2020MNRAS.494.2948P,2024ApJ...961..155K}. 

Due to the cannibalistic nature of this interaction, these compact binary MSPs, with observed orbital periods $P_{\rm orb}\lesssim 1\,\rm d$ \citep{2012ApJ...755..180P,2012ApJ...760L..36R,2013ApJ...765..158K,2013ApJ...769..108B,2013IAUS..291..127R}, are also called spider pulsars. So far, about 60 spiders have been observed and studied in the Galaxy \citep[for instance, see ][]{2009Sci...325..848A,2012Sci...338.1314P,2012ApJ...755..180P,2012ApJ...760L..36R,2013ApJ...765..158K,2013ApJ...769..108B}. There are two main categories of spiders \citep{2013IAUS..291..127R} based on the masses of the companions to the pulsar (denoted as $M_{\rm c}$): redbacks (RBs), which have companions with masses $0.1{\,\Msun}\lesssim M_{\rm c}\lesssim0.5\,\Msun$, and black widows (BWs), whose companions have masses of $M_{\rm c}<0.1\,\Msun$. The orbital period distribution is similar for both types ($\lesssim$1\,d).

Following classical binary evolution, the observed orbital periods of all spiders have not been explained, as they fall short by about an order of magnitude \citep{2002ApJ...565.1107P}. Fast winds that leave a binary isotropically and without interacting with anything else have a widening effect on the binary orbit \citep{1997A&A...327..620S}. Therefore, the involvement of this energetic radio pulsar wind has been invoked to explain the orbital properties of recycled pulsars \citep{1992A&A...265...65E,2002ApJ...565.1107P,2014ApJ...786L...7B,2015ApJ...798...44B,2013ApJ...775...27C}. There are a few ideas that have been proposed to explain spider MSPs. \citet{2013ApJ...775...27C} proposed strong and weak irradiation producing RBs and BWs, respectively. Another group studied the formation of spiders by including X-ray irradiation feedback, and they found that in order to reproduce the orbital parameters of spiders, cyclic mass transfer due to the effect of X-rays is required \citep{2014ApJ...786L...7B,2015ApJ...798...44B}. \citet{2020MNRAS.495.3656G} and \citet{2021MNRAS.500.1592G} argue that sustained magnetic braking (which couples to stellar wind and leads to angular momentum loss) is needed to reproduce spiders and that ablation alone is not sufficient. Some spiders are also thought to come from ultra-compact X-ray binaries (UCXBs), which are a sub-category of LMXBs with hydrogen-deficient companions and orbital periods of less than an hour \citep[see review by][]{2010NewAR..54...87N}. 

Since spider binaries have passed through a long mass-transfer phase, it has been suggested that they host some of the most massive \citep{2016arXiv160501665A,2018ApJ...859...54L, 2019ApJ...872...42S,2020mbhe.confE..23L} and fastest spinning NSs. The fastest spinning pulsar known so far spins at a frequency of 716\,Hz \citep[PSR\,J1748-2446ad spinning at about 1.4\,ms;][]{2006Sci...311.1901H} in the globular cluster Terzan\,5. Various reasons have been discussed to explain the lack of pulsars spinning faster than 1\,ms, for example an additional torque (from gravitational wave emission) that acts at high frequencies and spins down the pulsar \citep{1978MNRAS.184..501P,1984ApJ...278..345W,1998ApJ...501L..89B,2017ApJ...835....4B}. Since pulsar spins depend on the accretion phase, in this work we investigate the fastest spins possible for pulsars. Despite relatively large mass uncertainties, NSs in spider binaries tend toward higher masses \citep[median $M_{\rm NS}\sim 1.8\,\Msun$;][]{2020mbhe.confE..23L} in comparison to other systems, such as double NSs or wide NS plus white dwarf (WD) binaries \citep{2010Natur.467.1081D,2013Sci...340..448A,2020mbhe.confE..23L}.

In this study, we carry out detailed binary evolution simulations, calculating the evolution of the pulsar spin and taking into account isolated and mass-transfer evolution. The main aim is to investigate the effect of various assumptions of binary interaction on the pulsar spin evolution and to constrain them when comparing with observed spider properties. In Section\,\ref{sec:num_methods}, we describe the initial parameters of the simulated tracks, the added physics of pulsar spin evolution, and the treatment of the pulsar wind irradiation. Section\,\ref{sec:results} presents our simulated grids in the context of the observed population of spiders. In Section\,\ref{sec:discussion}, we compare our results with those in the literature and discuss the caveats involved. This is followed by the conclusions of our study in Section\,\ref{sec:conclusions}.

\section{Methods}
\label{sec:num_methods}
\subsection{Initial pulsar properties and binary physics}\label{sec:methods:ini_prop}
We use the detailed stellar evolution code Modules for Experiments in Stellar Astrophysics
\citep[\mesa;][]{2011ApJS..192....3P,2013ApJS..208....4P,2015ApJS..220...15P,2018ApJS..234...34P,2019ApJS..243...10P} to carry out our calculations for the evolution of spiders and included pulsar spin evolution. One of the benefits of using a detailed stellar code like \mesa\, is that compared to rapid population synthesis codes, detailed codes provide a more self-consistent calculation of the mass-transfer phase along with a physically motivated estimation of the stability of binary interaction. All the initial binaries in this study have a low- to intermediate-mass main-sequence star (MS; range of 1.0 to 4.5\,\Msun) and a point mass NS (range of 1.3 to 2.0\,\Msun). The initial orbital periods range from 0.6 to 4\,d. The simulations are run at Solar metallicity \citep[0.0142; ][]{2009ARA&A..47..481A}. Since most of the observed spider binaries with dynamical mass measurements are within the Milky Way \citep{2020mbhe.confE..23L}, we are focusing on Galactic spiders for our study. The physics in the \mesa\, grids follows the description in \citet{2023ApJS..264...45F}. We treat our numerically failed runs similar to \citet{2022arXiv220205892F}, by identifying failed binaries and rerunning them with changed parameters that will help computational stability and reach an acceptable termination condition. We discuss our treatment of the numerically failed binaries in Appendix\,\ref{sec:appendix:num_stab}.

\subsection{Isolated pulsar phase}\label{sec:methods:iso_pulsar}

As initial pulsar spin parameters, we take a spin of $1\,\rm s$ and a spin period derivative of $10^{-15}\,\rm s/s$. The corresponding magnetic field strength is $10^{12}\,\rm G$. These values correspond to the average spin parameters of observed young pulsars. We model the pulsar spin evolution by identifying different stages of the evolution and applying the corresponding equations, as described in the following sections. Pulsars in detached binaries, where the binary components are orbiting each other without transferring mass, are essentially assumed to spin down as isolated pulsars as their spins and magnetic fields are not affected by the companion. Since pulsars are rapidly spinning dipoles, they lose energy as they spin down in the form of magnetic dipole radiation, provided the dipole has an inclination. Hence, the spin-down of a pulsar results in the following decrease in its angular frequency ($\dot{\Omega}$):

\begin{equation}\label{eq:iso_evol}
    \dot{\Omega} = -\frac{2\mu^{2}\sin^{2}{\alpha}}{3\mu_{0}c^{2}I}\Omega^{3},
\end{equation}
where $\mu$ is the pulsar magnetic moment ($\mu \approx BR^{3}$), $\alpha$ is the angle between the rotation axis and magnetic axis, $\mu_{0}$ is the permeability of free space (it is unity in CGS units), $I$ is the pulsar moment of inertia (approximated as $2M_{\rm NS}R_{\rm NS}^{2}/5$ for a solid sphere, where $M_{\rm NS}$ and $R_{\rm NS}$ are the NS mass and radius), and $c$ is the speed of light. We assumed $\alpha=45\degree$, which is a simplistic approximation, as the inclination would decay with time \citep{2006ApJ...648L..51S,2017MNRAS.467.3493J}. The full range of $\alpha$ values would give an order of magnitude difference in $\dot{\Omega}$, at most \citep{2006ApJ...648L..51S}; hence, we did not vary this parameter in our study and fixed it at an intermediate value.

We evolved the isolated pulsar surface magnetic field ($B$) decay following \citet{2020MNRAS.494.1587C}: 
\begin{equation}\label{eq:B_isolated}
    B = (B^{i} - B_{\rm min})\times\exp{(-t/\tau_{\rm d})} + B_{\rm min},
\end{equation}
where $B^{i}$ is the initial surface magnetic field, $B_{\rm min}$ is the minimum magnetic field \citep[we take $B_{\rm min}=10^8\,\rm G$;][]{2006MNRAS.366..137Z,2011MNRAS.413..461O}, $t$ is time, and $\tau_{\rm d}$ is the magnetic field decay timescale \citep[we take $\tau_{\rm d}=3\,\rm Gyr$;][]{2019ApJ...877..122Y}. The superscripts $i$ and $f$ stand for initial and final values. Previous studies have explored $\tau_{\rm d}$ values ranging from a few Myr to a few Gyr, with decay timescales \devi{$\gtrsim 1\,\rm\,Gyr$} being \devi{favored} \citep{2008MNRAS.388..393K, 2019ApJ...877..122Y, 2020MNRAS.494.1587C}.
 
\subsection{Interacting binary phase}\label{sec:methods:interact_pulsar}

Once the companion to the pulsar fills its Roche lobe and the mass-transfer phase begins, accreted matter will carry angular momentum and spin up the pulsar. We consider mass accretion when there is a persistent accretion disk and use the criteria by \citet{1999MNRAS.303..139D} that describes the limiting mass-transfer rate below which thermal-viscous instabilities would prevent accretion. This limit is as follows:
\begin{equation}\label{eq:dubus_1999}
    \dot{M}_{\rm crit} = 3.2\times 10^{-9} \bigg(\frac{M_{\rm NS}}{1.4\,\Msun}\bigg)^{1/2} \bigg(\frac{M_{\rm c}}{1\,\Msun}\bigg)^{-1/5} \bigg( \frac{P_{\rm orb}}{1\,\rm d} \bigg)^{7/5} {\Msun\,yr^{-1}}.
\end{equation}

The change in the NS mass ($\dot{M}_{\rm NS}$) is determined by the efficiency of accretion ($\epsilon$), which is assumed to be equal to $1-\beta$, where $\beta$ is the fraction of mass transferred by the companion that is lost from the vicinity of the NS as a fast isotropic wind, taking away angular momentum from the NS. The rest of the material ($1-\beta$) is accreted by the NS, until the mass-transfer rate reaches the Eddington limit, beyond which all accretion is limited. Since the term $\beta$ affects the amount of accretion, lower values of $\beta$ result in a faster spin-up of the pulsar compared to higher values. Additionally, the value of $\beta$ affects the orbital evolution following the equation for the relative change in orbital separation \citep[$\dot{a}/a$;][]{2006csxs.book..623T}, assuming other changes in orbital angular momentum are represented by $\dot{J}_{\rm orb}$:
\begin{equation}
        \frac{\dot{a}}{a} = \frac{\dot{J}_\mathrm{orb}}{J_\mathrm{orb}} -2\,\bigg(1 + (\beta - 1) \frac{M_{\rm c}}{M_{\rm NS}} - \frac{\beta M_{\rm c}}{2 (M_{\rm c} + M_{\rm NS})}  \bigg) \frac{\dot{M}_{\rm c}}{M_{\rm c}}, 
\end{equation}
where, $M_{\rm c}$ and $M_{\rm NS}$ are companion and NS masses, respectively. Other processes to determine $\dot{J}_\mathrm{orb}$ (apart from mass transfer/loss) are magnetic braking \citep{1983ApJ...275..713R}, gravitational wave radiation \citep{1964PhRv..136.1224P}, and spin-orbit coupling. 

Theoretical and observational studies of various types of binaries have shown that the parameter $\beta$ could be more than 0.5 in LMXBs despite the evolution being mostly sub-Eddington \citep{1999A&A...350..928T,2001ASPC..229..423R,2002ApJ...565.1107P, 2005ApJ...629L.113J,  2011A&A...527A..83Z, 2012MNRAS.423.3316A, 2012MNRAS.425.1601T}. Some possible reasons for this are the propeller effect \citep{1975A&A....39..185I} and thermal-viscous instabilities in the accretion disk \citep{1981ARA&A..19..137P,1993Ap&SS.210...83M,1996ApJ...464L.139V}; all effects that would prevent accretion. Studies of accreting pulsars use values in the range of 0.5 to 0.7 \citep{2002ApJ...565.1107P, 2013ApJ...775...27C, 2014ApJ...791..127J, 2015ApJ...814...74J, 2020A&A...642A.174M, 2021MNRAS.506.4654W}. We add a prescription for thermal-viscous accretion disk instabilities (explained further below in Section\,\ref{sec:methods:interact_pulsar}) and explore three values of $\beta$ (0.0, 0.3, and 0.7) to study its effect on the pulsar spin evolution. This fractional value $\beta$ is also a proxy for the change in angular momentum when mass is accreted. We include $\beta$ even though we have a separate treatment for the propeller phase (discussed below in Section\,\ref{sec:methods:propel_regime}) and the aforementioned disk instabilities, since these have large uncertainties and there might still be inefficient transfer of angular momentum due to some unforeseen physical process.

The accreted matter suppresses the magnetic field due to rapid Ohmic decay \citep{1997MNRAS.284..311K, 1999MNRAS.303..588K, 1999MNRAS.308..795K}, which we include as an exponential decay with accreted mass \citep{2011MNRAS.413..461O, 2020MNRAS.494.1587C}, as follows:
\begin{equation}\label{eq:B_accretion}
    B = (B^{i} - B_{\rm min})\times\exp{(-\Delta M_{\rm NS}/M_{\rm d})} + B_{\rm min}\,.
\end{equation}
In the above equation, $\Delta M_{\rm NS}$ is the accreted matter and $M_{\rm d}$ is the magnetic field mass decay scale \citep[0.025, although other values for $M_{\rm d}$ have been explored in the range of 0.0033 to 0.05;][]{2011MNRAS.413..461O,2020MNRAS.494.1587C}. The accreted matter also causes a change in the angular momentum of the NS, which can be estimated as
\begin{equation}\label{eq:am_accreted}
    \dot{J} = \kappa \dot{M}_{\rm NS} R^{2}_{\rm NS} \omega_{\rm diff},
\end{equation}
where $\kappa$ is an efficiency factor and $\omega_{\rm diff}$ is the angular velocity of the incoming material \citep{2008MNRAS.388..393K, 2020MNRAS.494.1587C} that is defined by
\begin{equation}\label{eq:omega_diff}
    \omega_{\rm diff} = \sqrt{\frac{GM_{\rm NS}}{r^{3}_{\rm mag}}} - \sqrt{\frac{GM_{\rm NS}}{r^{3}_{\rm cor}}}.
\end{equation}
In the above equation, $r_{\rm cor}$ is the co-rotation radius, which is defined as $r_{\rm cor}=(GM_{\rm NS}/\Omega^{2})^{1/3}$ and $r_{\rm mag}$ is the magnetospheric radius, which is defined as follows \citep{2008MNRAS.388..393K,2020MNRAS.494.1587C}:
\begin{equation}\label{eq:r_mag}
    r_{\rm mag} = \frac{r_{\rm Alfv\acute{e}n}}{2} = \Bigg(\frac{2\pi^{2}}{G\mu^{2}_{0}}\Bigg)^{1/7} \Bigg(\frac{R^{6}_{\rm NS}}{\dot{M}_{\rm NS}M^{1/2}_{\rm NS}}\Bigg)^{2/7} B^{4/7},
\end{equation}
where, $r_{\rm Alfv\acute{e}n}$ is the ${\rm Alfv\acute{e}n}$ radius and $G$ is the gravitational constant. The two radii, $r_{\rm cor}$ and $r_{\rm mag}$, together with the light-cylinder radius, defined as $r_{\rm lc}=c/\Omega$, are associated with the NS and used to study the interacting binary phase. Using the three characteristic radii ($r_{\rm mag}$, $r_{\rm cor}$, and $r_{\rm lc}$), we define below the three main regimes of accretion:

\subsubsection{Accretion regime, $r_{\rm mag}<(r_{\rm cor},r_{\rm lc})$}\label{sec:methods:acc_regime}
If the mass-transfer rate is high enough that the magnetospheric radius is below the co-rotation radius (following Equation\,\ref{eq:r_mag}), we assume that accretion can proceed unimpeded. Since the magnetic field is weak, we assume that the inner accretion disk extends down to the NS surface. Hence, in Equation\,\ref{eq:am_accreted}, $R_{\rm NS}$ is equal to the NS radius, which we take as 12.5\,km \citep[for example,][]{2021ApJ...918L..29R}. Since, the increase in angular momentum comes from accreted material, we take $\dot{M}_{\rm NS}=\epsilon\dot{M}_{\rm c}=(1-\beta)\dot{M}_{\rm c}$, $\epsilon$ being the accretion efficiency. We also assume a $\kappa$ of unity, similar to \citet{2020MNRAS.494.1587C}.

\subsubsection{Propeller regime, $r_{\rm cor} \leq r_{\rm mag} < r_{\rm lc}$}\label{sec:methods:propel_regime}

While no mass is thought to have been accreted during the propeller phase \citep{1975A&A....39..185I}, it is uncertain how much material is propelled away. PSR J1023+0038 appears to be undergoing spin-down in a weak-propeller state, while still showing pulsed X-rays that signify accretion \citep{2015MNRAS.449L..26P,2015ApJ...807...62A,2018MNRAS.479L..12E}. Following Equation\,\ref{eq:omega_diff}, the change in angular momentum during the propeller phase is negative and the result is a spin down of the pulsar. We assume 10\% of mass is accreted out of the mass transferred by the companion ($\beta = 0.9$), while the rest is driven away from the NS, slowing it down. To account for spin down due to matter propelled away from the pulsar, we take $\dot{M}_{\rm NS}=\beta\dot{M}_{\rm c}$ in Equation\,\ref{eq:am_accreted}. Additionally, we assume a $\kappa$ of 0.001 in the Equation\,\ref{eq:am_accreted}, due to the uncertain nature of this phase.

\subsubsection{Ejection regime and pulsar wind phase}\label{sec:methods:eject_regime}
If the magnetic field is strong enough or the mass-accretion rate is low enough, the magnetosphere extends far outside of the light cylinder radius. Any accretion or transfer of angular momentum is prevented and the pulsar spin (and the corresponding pulsar magnetic field) is evolved as that of an isolated pulsar (see Section\,\ref{sec:methods:iso_pulsar}). We use two criteria to determine when the ejection regime and the consequent pulsar wind phase begins. First, the pulsar has accreted enough matter to spin up to MSP-like spins (we use 30\,ms as the identifying upper limit). The second condition for the onset of the pulsar wind phase is that the mass-transfer rate has decreased below the criteria for thermal-viscous instabilities and can no longer sustain a persistent accretion disk (see Equation\,\ref{eq:dubus_1999}), under the effect of X-ray irradiation during the RLO phase.

Once the ejection phase begins and as the pulsar spins down, the pulsar emits an energetic wind with a certain luminosity, known as the spin-down luminosity, which is defined using the spin properties of the pulsar as follows:
\begin{equation}\label{eq:spindown_L}
    L_{\rm pulsar} = \frac{4\pi^{2} I \dot{P}_{\rm spin}}{{P}_{\rm spin}^{3}}.
\end{equation}
\devi{As this energetic wind hits the companion, it ablates the outer surface of the star, which, when observed, signifies the presence of a spider pulsar \citep{1988Natur.334..225K, 1989ApJ...343..292R, 1989ApJ...336..507R, 1996ApJ...473L.119S}.} In the above equation, the initial values of the NS spin ($P_{\rm spin}$) and change in NS spin with time ($\dot{P}_{\rm spin}$) are determined by the preceding accretion phase (see Section\,\ref{sec:methods:acc_regime}) and follow the spin evolution of a detached pulsar (see Section\,\ref{sec:methods:iso_pulsar}). The irradiation from the pulsar wind drives a mass loss from the companion surface ($\dot{M}_{\rm c, irr}$) defined by \citet{1992MNRAS.254P..19S}:
\begin{equation}\label{eq:mdot_irrad}
    \dot{M}_{\rm c, irr} = -f_{\rm pulsar}\times\frac{L_{\rm pulsar}}{2v^{2}_{\rm esc}}\bigg(\frac{R_{\rm c}}{a}\bigg)^{2},
\end{equation}
where $f_{\rm pulsar}$ is the efficiency that describes the transfer of energy from the incoming spin-down luminosity to the generated wind loss. It also stands for any geometrical beaming effects that might be present during the process. The other parameters in the above equation include $L_{\rm pulsar}$ as the spin-down luminosity, $v_{\rm esc}$ as the escape velocity from the companion star surface, $R_{\rm c}$ as the companion radius, and $a$ as the binary separation. \devi{The effect of this increased mass loss from the companion is described by the fast mode or Jean's mode of mass transfer \citep{1924MNRAS..85....2J,1997A&A...327..620S}. This mode of mass transfer assumes that a small fraction of the stellar mass is lost in the form of fast, spherically symmetric winds that leave the binary without interacting with any binary components, escaping with the specific angular momentum of the mass-losing star. The orbital angular momentum ($J_{\rm orb}$) is defined as}
\devi{\begin{equation}
    J_{\rm orb} = \mu \sqrt{Ga(M_{\rm c}+M_{\rm NS})},
\end{equation}}
\devi{where $\mu$ is the reduced mass of the binary. The change in orbital angular momentum per unit of reduced mass ($J/\mu$) is negligible and the orbit expands accommodate the change in total mass \citep{1963ApJ...138..471H}. As the effect of irradiation will not necessarily be spherically symmetric, we have the efficiency term $f_{\rm pulsar}$ as described above \citep{1992MNRAS.254P..19S,2013ApJ...775...27C}.}

The definition of the three regimes of mass transfer described above based on the relative radii of the magnetosphere, co-rotation radius, and light cylinder radius, are only approximate estimations. There is work showing that the boundaries between these regimes are bit more complicated \citep{2015arXiv150403996E, 2018MNRAS.479L..12E}. However, since they are highly uncertain, we assume the more straightforward definitions described.

\subsection{Observations used for comparisons to simulations}
\label{sec:obs}
To compare our simulations to observations we use a \devi{catalog} of BWs, RBs, and huntsmen (including candidates). A total of 57 spiders are used, 52 of which are \devi{cataloged} in \citet[][and references therein]{2021JCAP...02..030L}. The more recent ones are from various other works in the literature, and an updated \devi{catalog} has been prepared by \citet[][and references therein]{Nedreaas_master_thesis}. The reported values for the companion masses are in terms of the minimum mass (denoted as $M_{\rm c,min}$), calculated assuming a pulsar mass of $1.4\,\Msun$ and inclination of $90\degree$ from observations of radio pulses. Only for one of the huntsmen source in the \devi{catalog} have radio pulses been confirmed and $M_{\rm c,min}$ estimated \citep[1FGL\,J1417.7--440][]{2016ApJ...820....6C}. While the other huntsmen, 2FGL\,J0846.0+2820, only has Fermi Large Area Telescope (LAT) observations \citep{2017ApJ...851...31S}.

\section{Results}
\label{sec:results}

\begin{figure*}[!ht]
\centering
\includegraphics[width=\linewidth]{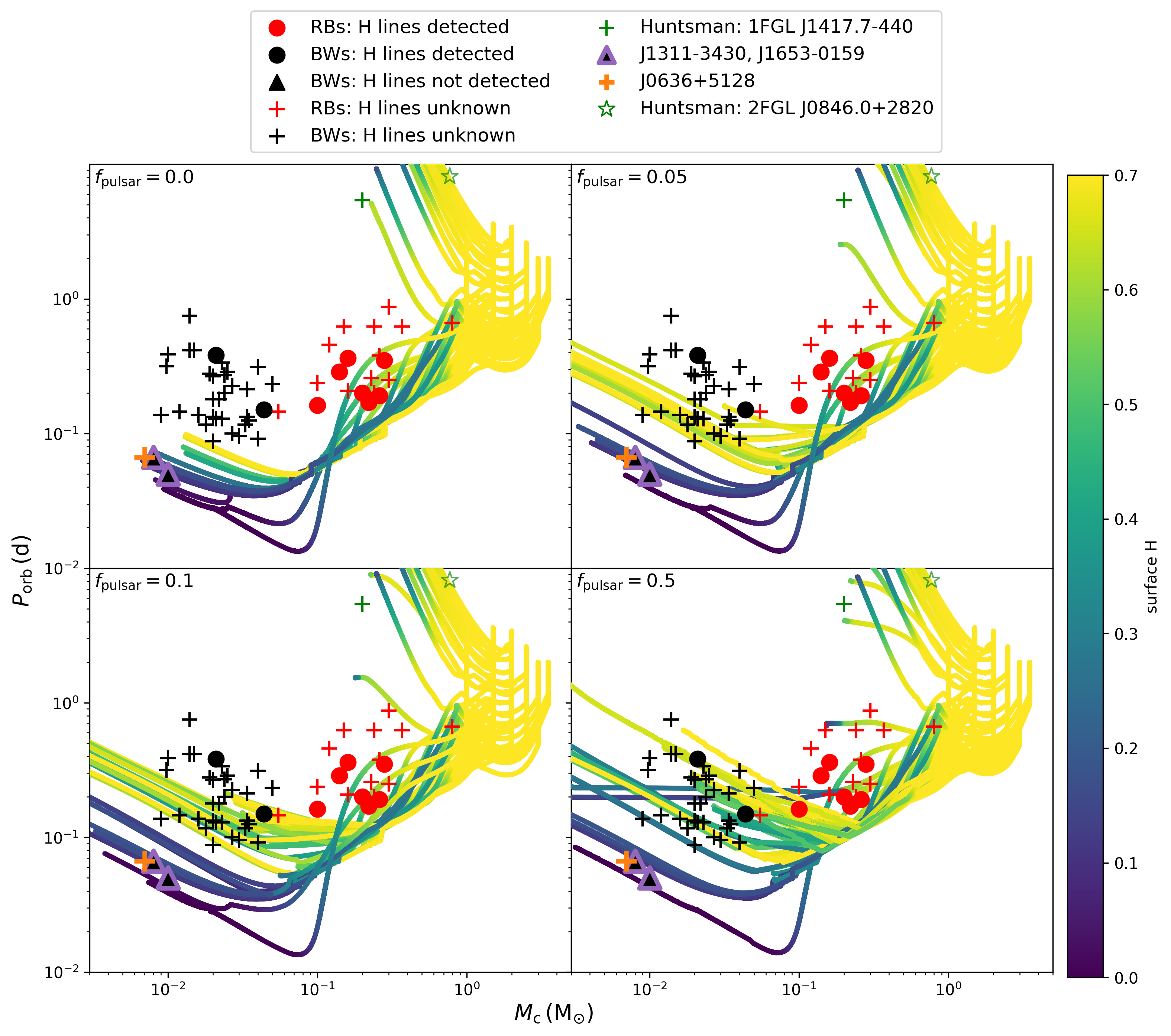}
\caption{Orbital period evolution versus companion mass of the simulated binaries for $\beta=0.3$ with an initial NS of mass $1.3\,\Msun$ and $f_{\rm pulsar}$ values of 0.0 (top left), 0.05 (top right), 0.1 (bottom left), and 0.5 (bottom right). The term $M_{\rm c}$ is the mass of the companion to the NS, and $P_{\rm orb}$ is the orbital period. The color bar shows the changing surface abundance of the companion. We also compare our tracks to estimated $M_{\rm c,min}$ and observed $P_{\rm orb}$ for BWs (black crosses) and RBs (red crosses), and huntsmen spiders (green crosses), including whether H-lines have been detected (filled circles) or not (filled triangles) or if it is unknown (crosses), from \citet[][and references therein]{Nedreaas_master_thesis}. The green star is the huntsman candidate 2FGL\,J0846.0+2820, observed by Fermi-LAT \citep{2017ApJ...851...31S} (surface H is unknown). The three tidarrens (J1311-3430, J1653-0159, and J0636+5128) are also shown. The initial binaries have a MS star (range of 1.0 to 4.5\,\Msun) and initial periods of 0.6 to 4.0\,d.}
\label{fig:beta03}
\end{figure*}

\begin{figure}[!ht]
\centering
\includegraphics[width=\linewidth]{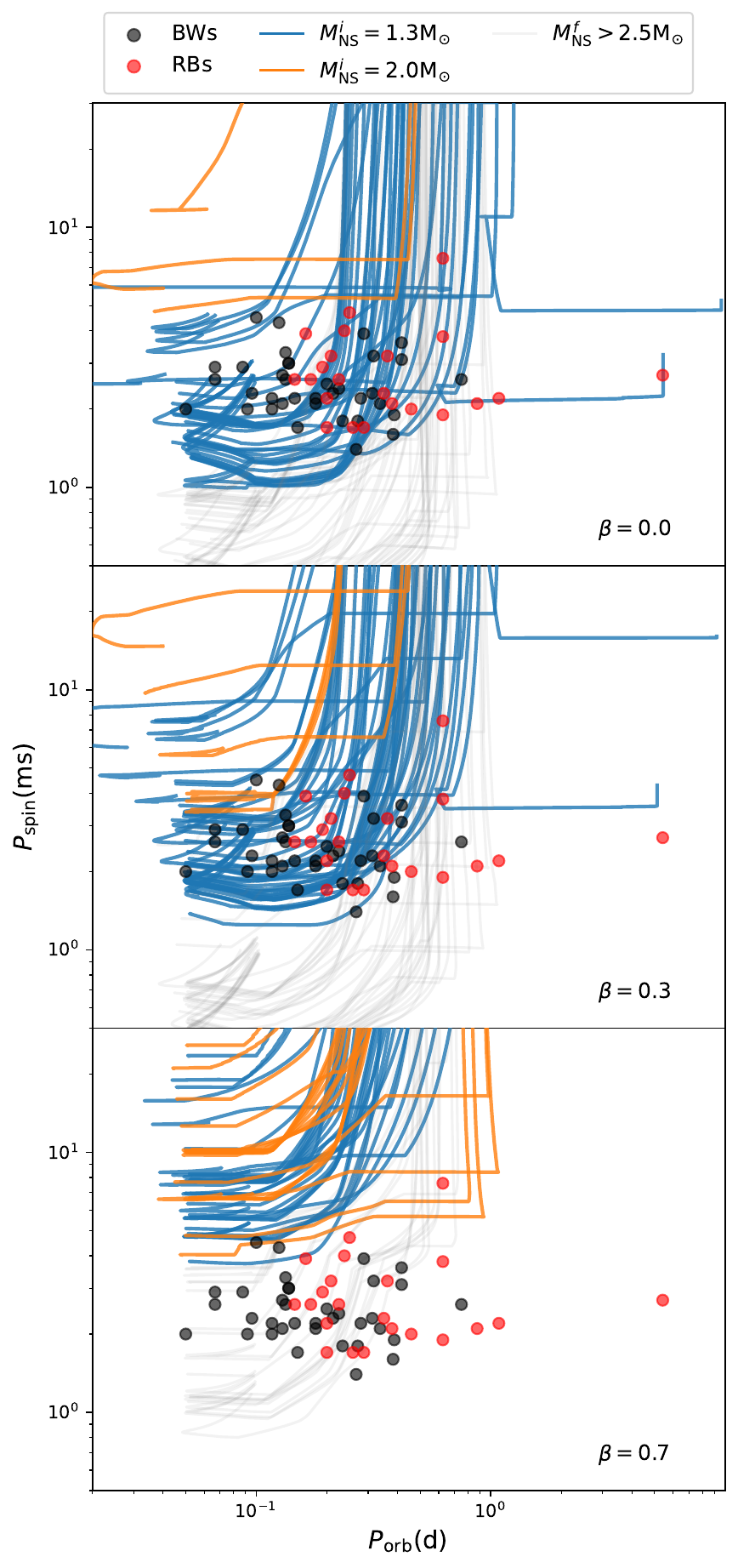}
\caption{ \devi{Evolution of the NS spins ($P_{\rm spin}$) versus the orbital periods ($P_{\rm orb}$) in converging binaries for $\beta=0.0$ (top), $\beta=0.3$ (middle), and $\beta=0.7$ (bottom). We distinguish the binaries if the final NS mass was above $2.5\,\Msun$ (faint gray curves) or below it, which are further grouped by two initial NS masses ($M^{i}_{\rm NS}$) 1.3\,\Msun (blue curves) and 2.0\,\Msun (orange curves). We also compare to the observed spins of BWs and RBs, shown by the black and red circles, respectively.}  }
\label{fig:spin_hist}
\end{figure}

\subsection{Mass-accretion efficiency}
\label{sec:results:acc_eff}

We investigate the formation of different branches of compact MSP binaries using detailed evolutionary sequences. Figures\,\ref{fig:beta03} and \ref{fig:beta07} show the evolutionary tracks with $\beta$ of 0.3 and 0.7, respectively, each with different values of $f_{\rm pulsar}$. The tracks correspond to initial NS masses of 1.3\,\Msun and have final NS spins $\lesssim\rm\,30\,ms$. While $\beta=0.7$ reproduces the very compact spiders, it cannot explain a large part of the binary parameter space where these systems are observed. This behavior is irrespective of the pulsar wind efficiencies. Similar behavior is seen for huntsmen spiders ($P_{\rm orb}$ > 1.0\,d), as $\beta=0.7$ does not reproduce the small population of observed huntsmen spiders. With $\beta$ of 0.3, a large part of the observed parameter space is explained and on increasing $f_{\rm pulsar}$, more and more BWs and RBs' observed orbital properties are reproduced. The reason for the difference between $\beta=0.3$ and $\beta=0.7$ can be understood by looking at Equation\,\ref{eq:mdot_irrad}, which shows that the mass lost due to pulsar wind irradiation is directly proportional to the pulsar spin-down luminosity and in turn depends on the pulsar spin at the end of the RLO phase (Equation\,\ref{eq:spindown_L}). Since the preceding accretion phase dictates the spin-up of the pulsar, we can compare the resulting pulsar spins in our grids with observations to see why more accretion helps. 


We compared the observed BWs \devi{and RBs} to the pulsar spin evolution from the grid in Figure~\ref{fig:spin_hist}, corresponding to $f_{\rm pulsar}=0.0$ and $\beta=$ 0.0, 0.3, and 0.7. The figure shows the binaries \devi{for two} initial NS masses \devi{1.3\,\Msun and 2.0\,\Msun}, and also separates out the cases where the final NS mass exceeds 2.5\,\Msun, which is an approximate upper limit for the NS mass. The NSs with final masses above 2.5\,\Msun are considered to have collapsed into black holes (BHs) and would no longer be observed as radio pulsars. The value $\beta=0.0$ leads to a majority of the NSs \devi{with $M^{i}_{\rm NS}>1.3\,\Msun$} gaining spins in the sub-MSP range ($\lesssim 1$~ms) and having final NS masses $\gtrsim 2.5\,\Msun$. With a $\beta$ of $0.7$ (or an accretion efficiency of $0.3$), the fastest spinning Galactic BW observed so far \devi{can be reproduced} \citep[PSR\,J0952--0607, with a spin period of 1.41\,ms;][]{2017ApJ...846L..20B}. However, these short spins correspond to the case where the final NS mass exceeds the maximum NS mass limit. Removing these, the minimum spin period for $\beta=0.7$ is \devi{3.8\,ms}, which is \devi{almost thrice} the minimum observed BW spin. Looking at the case of $\beta=0.3$ in Figure\,\ref{fig:spin_hist}, \devi{most of} the observed spins can be reproduced. Comparing the minimum \devi{final} spin period for $\beta=0.3$ (1.38\,ms) and the corresponding final pulsar mass ($2.36\,\Msun$) with PSR\,J0952--0607, we find that our simulations can reproduce well this extreme BW pulsar \citep[its estimated pulsar mass is $2.35^{+0.17}_{-0.17}\,\Msun$;][]{2022ApJ...934L..17R}. 

\devi{In Figure\,\ref{fig:spin_hist},} there are some binary tracks that ended with final spins below 1\,ms, in the sub-MSP regime (all with $M^{i}_{\rm NS}>1.3\,\Msun$). The NSs with $M^{i}_{\rm NS}=1.3\,\Msun$ gained a maximum of $1.06\,\Msun$ and did not spin up to the sub-MSP regime (final spin was 1.37\,ms). All the binaries where the final NS mass exceeded the maximum NS mass limit, were spun up to sub-milliseconds. Hence, the mass required to spin up an NS to sub-millisecond range is high enough to collapse it into a BH. Additionally, the sub-millisecond regime is mainly dominated by initially more massive NSs (\devi{majority of the $M^{i}_{\rm NS}=2.0$\,\Msun tracks in Figure\,\ref{fig:spin_hist} appear to be above the maximum NS mass limit}) while the initially less massive NSs are not always able to accrete enough matter to enter the sub-millisecond regime. This is because more massive NSs leads to more a stable and longer lasting mass transfer phase \citep{2020A&A...642A.174M}. Comparing the final spin distributions between $\beta=0.3$ and $\beta=0.7$, we can see that $\beta=0.7$ results in slower spinning NS, which would then have lower spin-down luminosities as $L_{\rm pulsar} \propto P^{-3}_{\rm spin}$, and not affecting the orbital evolution as much. Increased wind mass loss from the companion would lead to increased orbital widening during the radio pulsar phase. \devi{With $\beta = 0.0$, while the entire observed spider parameter space is reproduced, the spin evolution extends down to 1\,ms. The value of $\beta$ that is most consistent with the observed minimum pulsar spin is $0.3$. Additionally, $\beta=0.3$ leaves some room for uncertainty in accretion and in our implementation of the X-ray irradiated accretion disk instabilities. With future observations of spiders, any spins approaching 1\,ms would signify more conservative nature of RLO at play. Hence, $\beta\lesssim 0.3$ is required to explain the observed spins of spider pulsars. }

\begin{figure*}[!ht]
\centering
\includegraphics[width=\linewidth]{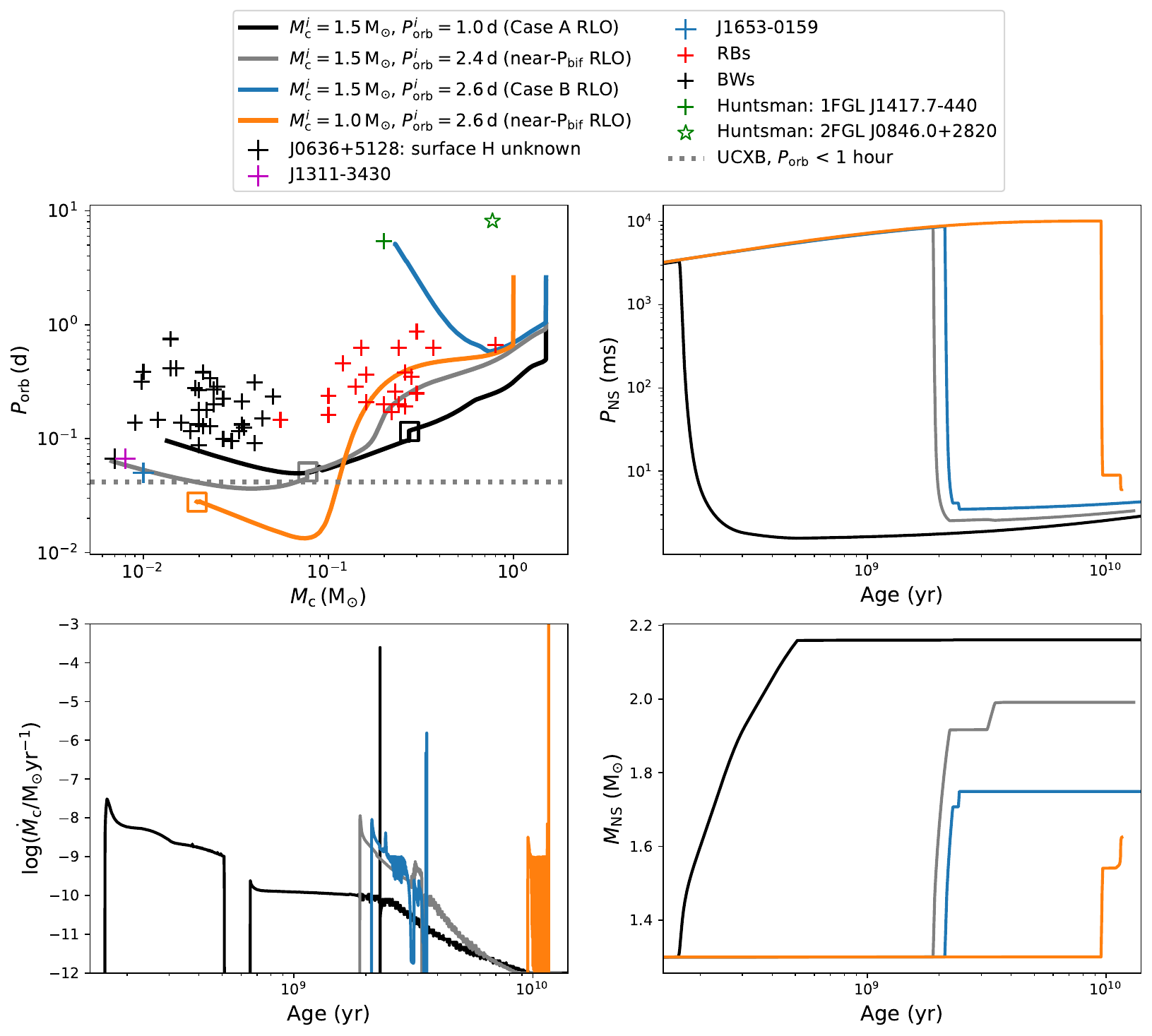}
\caption{Evolution of selected binaries from our simulated grids for $\beta=0.3$, $f_{\rm pular}=0.0$, and $M^{i}_{\rm NS}=1.3\,\Msun$. The three different kinds of binaries that we discuss in the text are studied here: case A ($P^{i}_{\rm orb}=1.0\rm\,d$), near-$P_{\rm bif}$ ($P^{i}_{\rm orb}=2.4\rm\,d$), and \devi{case B or }diverging ($P^{i}_{\rm orb}=2.6\rm\,d$). Also shown is the binary that goes down to the orbital period of 19\,min (orange curve) for comparison, which is the minimum in the grid with $f_{\rm pulsar}=0.0$. \devi{Top left shows the orbital period versus companion mass, where} $M_{\rm c}$ is the mass of the companion to the NS and $P_{\rm orb}$ is the orbital period. The formation of a fully convective star \devi{and the start of the ejection phase are shown} as squares. We also compare our tracks to estimated $M_{\rm c,min}$ and observed $P_{\rm orb}$ for BWs (faint black crosses), RBs (faint red crosses), and huntsmen spiders (faint green crosses), from \citet[][and references therein]{Nedreaas_master_thesis}. The green star is the huntsman candidate 2FGL\,J0846.0+2820, observed by Fermi-LAT \citep{2017ApJ...851...31S}. The tidarrens, namely, J1311--3430, J1653--0159, and J0636+5128, are highlighted as pink, blue, and black crosses, respectively. \devi{Top right, bottom left, and bottom right show the time evolution of the NS spin, companion mass loss rate, and NS mass, respectively, for the same binaries. The time axis is zoomed in to focus on the spin-up phase.} }
\label{fig:ucxb_eg1}
\end{figure*}

\subsection{Orbital evolution and surface hydrogen abundance}
\label{sec:results:NONirrad_grid}

The top left panel in Figure\,\ref{fig:beta03} shows the evolutionary tracks with no pulsar wind irradiation ($f_{\rm pulsar}=0.0$). The color bar corresponds to the surface H abundance of the companion, which can be compared to optical spectroscopic observations of spiders. We mark in Figure\,\ref{fig:beta03}\devi{,} spiders where H lines have not been detected in their optical spectra (filled triangles) and those where H lines are detected (filled circles). Two of the most compact BWs, PSR\,J1311--3430 \citep{2012Sci...338.1314P,2012ApJ...760L..36R} and PSR\,J1653--0159 \citep{2014ApJ...794L..22K,2014ApJ...793L..20R}, with negligible surface H abundance, are easily reproduced by our evolutionary tracks. These sources also form a sub-category of BWs, called tidarrens, with $P_{\rm orb}\lesssim 2$\,hr and $M_{\rm c}\lesssim 0.015\,\Msun$ \citep{2016ApJ...833..138R}. They are reproduced regardless of the value of $f_{\rm pulsar}$. Their evolutionary tracks also pass through the ultra-compact X-ray binary region (UCXB), with orbital periods less than one hour, with the minimum orbital period being 30.8\,min. Most UCXBs are also considered to be H-deficient due to their small orbits \citep{1986ApJ...304..231N,2006MNRAS.370..255N}, similar to our binary evolution tracks for tidarrens. Hence, tidarrens (the most compact BWs with the lightest companions) passed through a UCXB phase in their previous evolution. Additionally, we also see these systems reproduced in Figure\,\ref{fig:beta07} (with $\beta=0.7$), since not a lot of accretion is needed to form these sources. 

We can understand their evolution better in Figure\,\ref{fig:ucxb_eg1} \devi{(top left panel)} where we show tracks that pass through the tidarren region (solid gray line, with an $M^{i}_{\rm c}=1.5\,\Msun$ and $P^{i}_{\rm orb}=$2.4\,d) and compare these with other tracks with initially narrower and wider orbits. All the tracks shown in the figure have an $M^{i}_{\rm NS}=1.3\,\Msun$. We also show a binary that goes well into the UCXB region during its evolution, to reach the overall minimum orbital period (19\,min) in the grid slice (with an $M^{i}_{\rm c}=1.0\,\Msun$ and $P^{i}_{\rm orb}=$2.6\,d). With increasing orbital periods, the binaries go from starting RLO with the companion on the MS (at $P_{\rm orb}=0.5$\,d, also called case A RLO), to starting RLO when the companion has exhausted core-H (also called case B RLO, at $P_{\rm orb}=1.1$\,d). \devi{We also see this delay in RLO in bottom right in Figure\,\ref{fig:ucxb_eg1}, showing the time evolution of the mass-loss rates of the companions.} This transition occurs at the bifurcation period ($P_{\rm bif}$), because companions in wider orbits have more time to evolve and move out of the MS. In the presented example, $P_{\rm bif}$ is between $P_{\rm orb}=0.93$\,d (with $P^{i}_{\rm orb}=2.4$\,d and $M^{i}_{\rm c}=1.5\,\Msun$) and $P_{\rm orb}=1.1$\,d (with $P^{i}_{\rm orb}=2.6$\,d and $M^{i}_{\rm c}=1.5\,\Msun$). We term systems just below $P_{\rm bif}$ as near-$P_{\rm bif}$ binaries; similar studies have shown that the formation of UCXBs from LMXBs happens just below $P_{\rm bif}$ \devi{\citep{2002ApJ...565.1107P,2003MNRAS.340.1214P,2019RAA....19..110H,2023ApJ...950...27G}}.

For binary evolution governed only by conservative mass transfer, the orbit would expand as long as the mass is being transferred from the less massive to the more massive star, while it would contract if the companion star is more massive. With the involvement of magnetic braking, this evolution is not as simple. In NS binaries with MS companions having masses around $1\,\Msun$, orbital angular momentum evolution is dominated by magnetic braking, which contracts the orbits. Since case A RLO proceeds on a nuclear timescale during the evolution of the stellar MS, any mass loss via winds is not fast enough to dominate over magnetic braking. \devi{We can see this nuclear timescale RLO phase in Figure\,\ref{fig:ucxb_eg1} (bottom left panel).} In compact binaries \devi{with orbital periods} $\lesssim 1\rm\,d$, angular momentum losses due to gravitational wave radiation are also \devi{significant} and further contract the orbit. Once the companion in case A binaries decreases to about 0.2 to 0.3\,\Msun, it develops a fully convective structure and magnetic braking stops operating \citep{1983ApJ...275..713R,1983A&A...124..267S,1989A&A...208...52P,2002ApJ...565.1107P}. This is most prominent for the binary in Figure\,\ref{fig:ucxb_eg1} \devi{(top left panel)} with $M^{i}_{\rm c}=1.5\,\Msun$ and $P^{i}_{\rm orb}=1$\,d; there is a decrease in the orbit without any change in mass in the companion mass. The companion detaches from its Roche lobe as the main process driving RLO (the first RLO phase for this binary) has stopped operating, until it fills its Roche lobe again due to gravitational wave radiation continuing to contract the orbit (which is the drop in orbital period seen in the figure). The gravitational wave radiation sustains this second RLO phase and the orbit eventually expands. \devi{Correspondingly, we also see this break in the mass-transfer rate in the same figure after which the mass-transfer rate increases again for the second RLO phase. The overall effect of this RLO phase is seen in the top and bottom left panels of Figure\,\ref{fig:ucxb_eg1}, case A NSs undergo accretion much more ($\sim 0.85\,\Msun$) and have an earlier spin-up than other binaries.}

For case B RLO  (corresponding to $M^{i}_{\rm c}=1.5$\,\Msun and $P^{i}_{\rm orb}=2.6$\,d in Figure\,\ref{fig:ucxb_eg1}), strong stellar winds take away stellar angular momentum. Since angular momentum lost due to stellar winds expands the orbits and the effect of magnetic braking is weaker in comparison, we get diverging orbits. \devi{The companion has almost an order of magnitude higher mass-loss rate than case A.} The binary is at the very end of the companion MS at the onset of RLO (central H abundance less than 0.1); hence, there is a small contraction in the orbit before it expands during the giant phase. The resulting binaries from these diverging systems are NS/He-WDs in wide orbits. \devi{Since this is a shorter-lived phase than case A, the NS does not accrete as much for the same initial donor mass ($\sim 0.4\,\Msun$).}

\devi{The mass-transfer phase for near-$P_{\rm bif}$ systems is similar to case B, though the outcomes differ. The companions in near-$P_{\rm bif}$ binaries have not fully left the MS at the start of RLO, central abundance is close to about 0.1 and magnetic braking is still operating. The stellar radius is about 1.5 times that of the case A companion at the onset of RLO, making the MS winds almost an order of magnitude stronger ($\sim 10^{-12}\,\Msun\rm\,yr^{-1}$).} The companion becomes fully convective at a significantly lower mass than case A systems ($M_{\rm c}\lesssim 0.1\,\Msun$, see gray and orange lines in the top left panel of Figure\,\ref{fig:ucxb_eg1}). The magnetic braking effect increases as the orbit shrinks \citep[$\propto$\,$P_{\rm orb}^{-3}$;][]{1983ApJ...275..713R}.\devi{ The reason for the fully convective structure forming after the star has lost most of its mass is the duration of the} first RLO phase. As the companion loses mass, its central temperature decreases due to loss in internal pressure. The case A binary has a RLO phase that started earlier \devi{and lasted longer} compared to the near-$P_{\rm bif}$ binary \devi{(seen also in Figure\,\ref{fig:ucxb_eg1}, bottom left panel)} providing enough time to cool internal structure of the star and making it fully convective while the companion is still on the MS. The near-$P_{\rm bif}$ binary has stellar winds that, along with the magnetic braking driven RLO, strip the outer layers of the star \devi{over a much shorter duration than case A}. The shorter the first RLO phase is, the further the orbit contracts into the UCXB region. 


The binary that reaches the minimum $P_{\rm orb}$ (19\,min; Figure\,\ref{fig:ucxb_eg1} \devi{top left panel}) in the grid slice (with an $M^{i}_{\rm c}=1.0\,\Msun$ and $P^{i}_{\rm orb}=2.6$\,d) forms a fully convective companion \devi{after the companion reaches $\lesssim0.02\,\Msun$. The NS present for this binary also accretes the least in our selected binaries, and the NS spin-up occurs the latest. Toward the end of the second RLO phase, while the companion star is detaching from its Roche lobe, orbital angular momentum losses due to gravitational radiation are still operating (strong for periods $\lesssim1\,$d). This causes the companion star to fill its Roche lobe again. As the companion is fully convective, it responds by expanding \citep{1987ApJ...318..794H} further extracting angular momentum from the orbit (due to spin-orbit coupling). As both angular momentum losses (due to RLO and spin-orbit coupling) reduce the binary separation, the Roche-lobe radius is proportionally affected and the companion overfills its Roche lobe, increasing the mass-transfer rate to $\lesssim 10^{-2}\,\Msun\rm\,yr^{-1}$ and triggering a dynamical instability.} 


There is another source in the same observed parameter space as the two tidarrens (see Figure\,\ref{fig:beta03}), for which the surface H abundance is unknown: PSR J0636+5128 \citep{2014ApJ...791...67S,2018ApJ...862L...6D}. Since it is in the same region of the parameter space as PSR\,J1311--3430 and PSR\,J1653--0159 and lies on similar evolutionary tracks, it is very likely this source had a similar formation history and will be H deficient on its surface. Since tidarrens can be produced with any value of $f_{\rm pulsar}$, the effect of irradiation on their evolution is not significant. They reach their extremely low masses due to the extended RLO phase along with stellar winds during the giant phase, during which they lose their entire H envelope. As we saw in Figure\,\ref{fig:ucxb_eg1}, the lower-right part of the RB population is also reproduced. These systems are the very same that pass through the UCXB region and lead to the tidarrens, and correspond to systems that start their RLO just below $P_{\rm bif}$. As there is no irradiation included in their evolution, these systems could depict some of the non-irradiated RBs observed. Since all currently observed BWs and about half of the currently observed RBs are strongly irradiated \citep{2023MNRAS.525.2565T}, we investigate higher values of $f_{\rm pulsar}$ to form RBs in the subsequent sections.

\begin{figure*}[!ht]
\centering
\includegraphics[width=\linewidth]{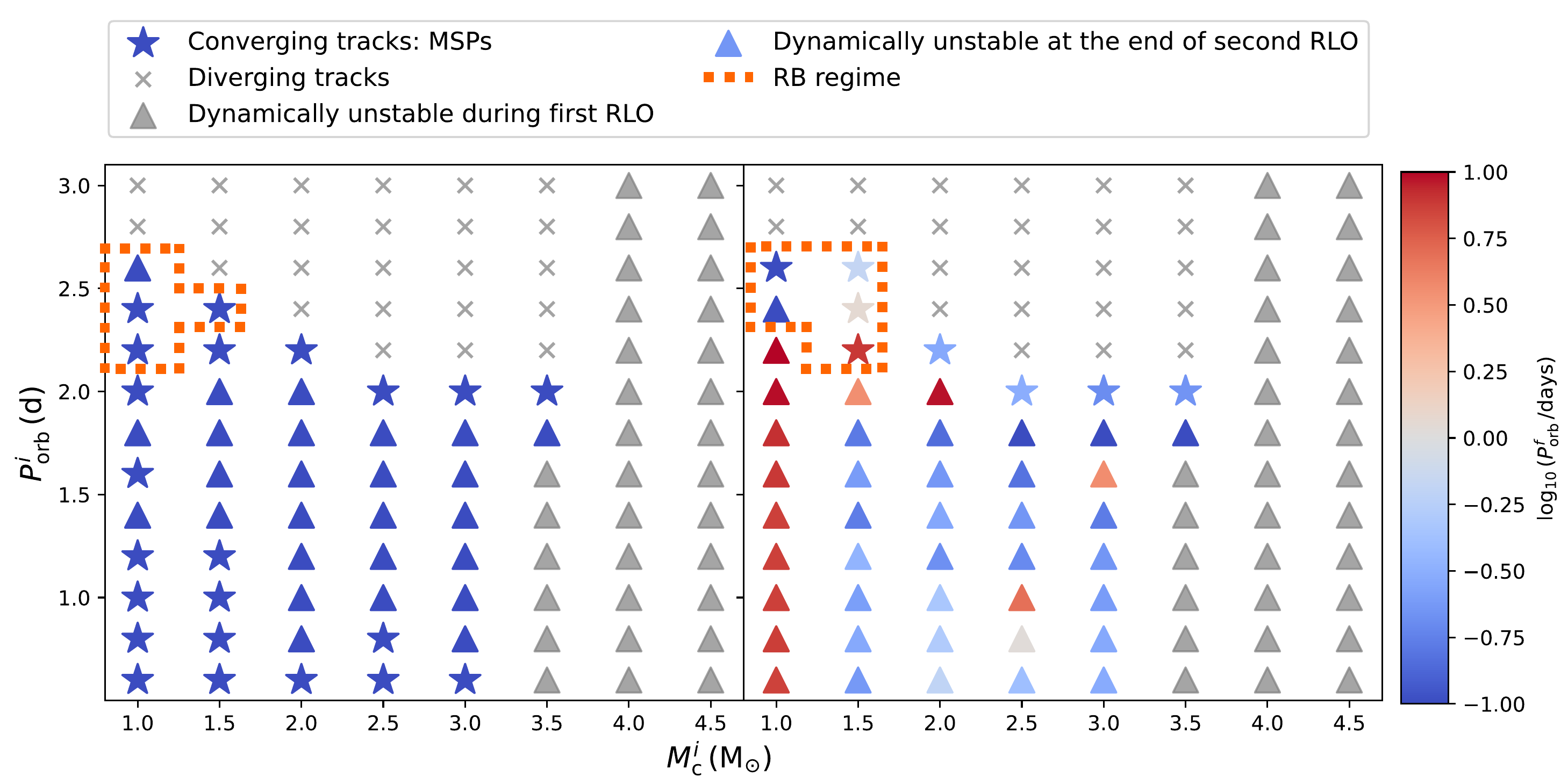}
\caption{Allowed initial parameter space to form spider MSPs for an initial NS of mass $1.3\,\Msun$ and $f_{\rm pulsar}$ values of 0.0 (left) and 0.5 (right). The term $M^{i}_{\rm c}$ is the initial mass of the companion to the NS, and $P^{i}_{\rm orb}$ is the initial orbital period. The color bar shows the final orbital periods ($P^{f}_{\rm orb}$) and the gray symbols did not attain MSP-like spins. We show the end states of the tracks, if they encountered a dynamical instability \devi{(with gray triangles or with colorful triangles depending on if onset of instability occurred during the first RLO phase or the second, respectively)} or ended with one of the criteria for successful termination (star symbols). The diverging symbols are shown by gray crosses. We also show the near-$P_{\rm bif}$ binaries that passed through the RB part of the observed parameter space enclosed by dashed orange lines. The magenta triangle shows the initial parameters of the binary that reproduces PSR\,J0952--0607 (with observed spin period of 1.41\,ms).}
\label{fig:BWParam_space_f1}
\end{figure*}

\begin{figure*}[!ht]
\centering
\includegraphics[width=\linewidth]{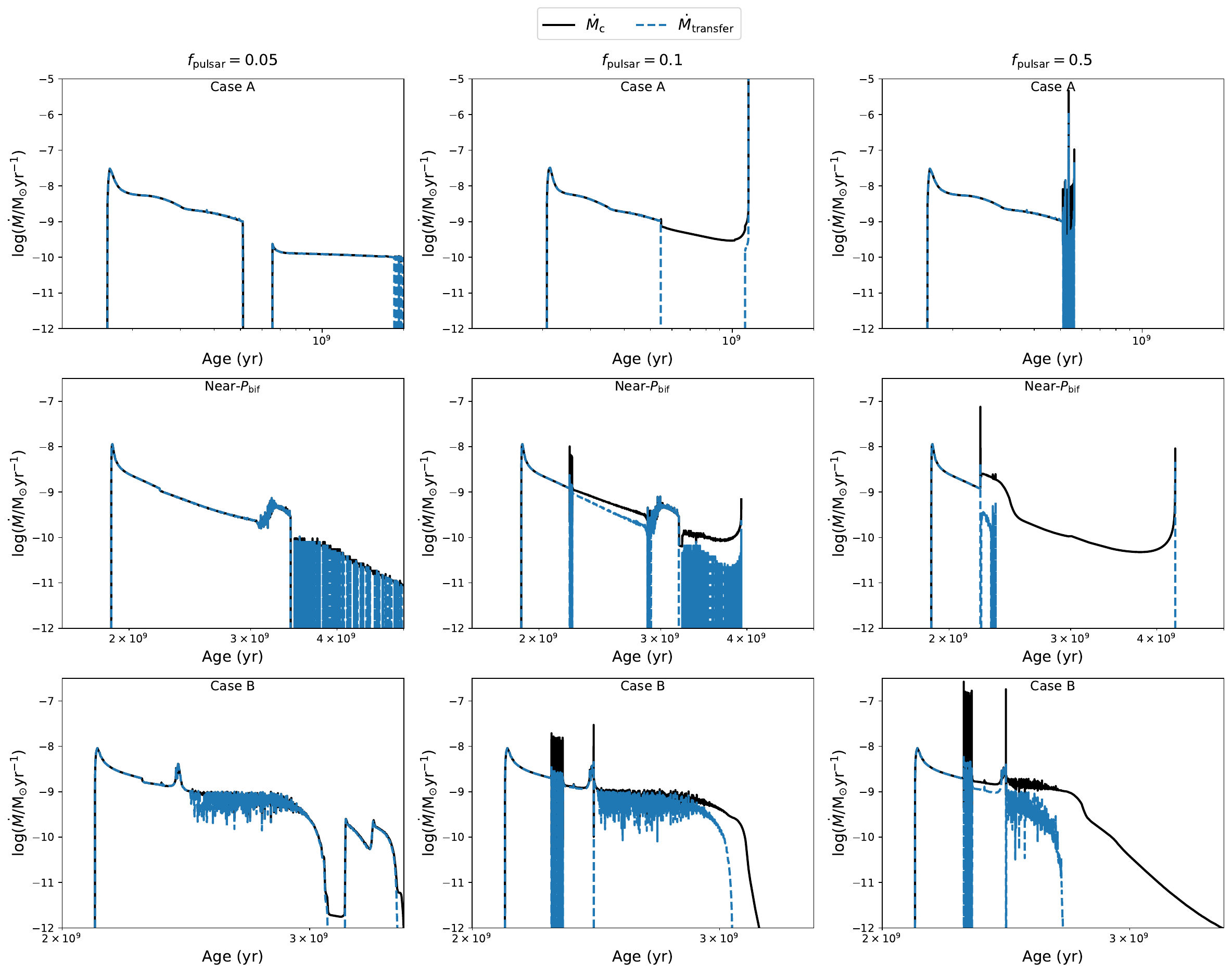}
\caption{\devi{Evolution of the mass-loss rates ($\dot{M}_{\rm c}$, black curves) and the mass-transfer rates ($\dot{M}_{\rm transfer}$, blue curves) of the companion star. The initial binary masses are $M^{i}_{\rm c}=1.0$\,\Msun and $M^{i}_{\rm NS}=1.3$\,\Msun, with $\beta=0.3$. We present the three cases of interacting binaries studied, case A (top row; with $P^{i}_{\rm orb}=1.0$\,d), near-$P_{\rm bif}$ (middle row; with $P^{i}_{\rm orb}=2.4$\,d), and case B (bottom row; with $P^{i}_{\rm orb}=2.6$\,d), each for three values of $f_{\rm pulsar}=0.05$ (left column), 0.1 (middle column), and 0.5 (right column).}}
\label{fig:bw_eg2}
\end{figure*}

\begin{figure*}[!ht]
\centering
\includegraphics[width=\linewidth]{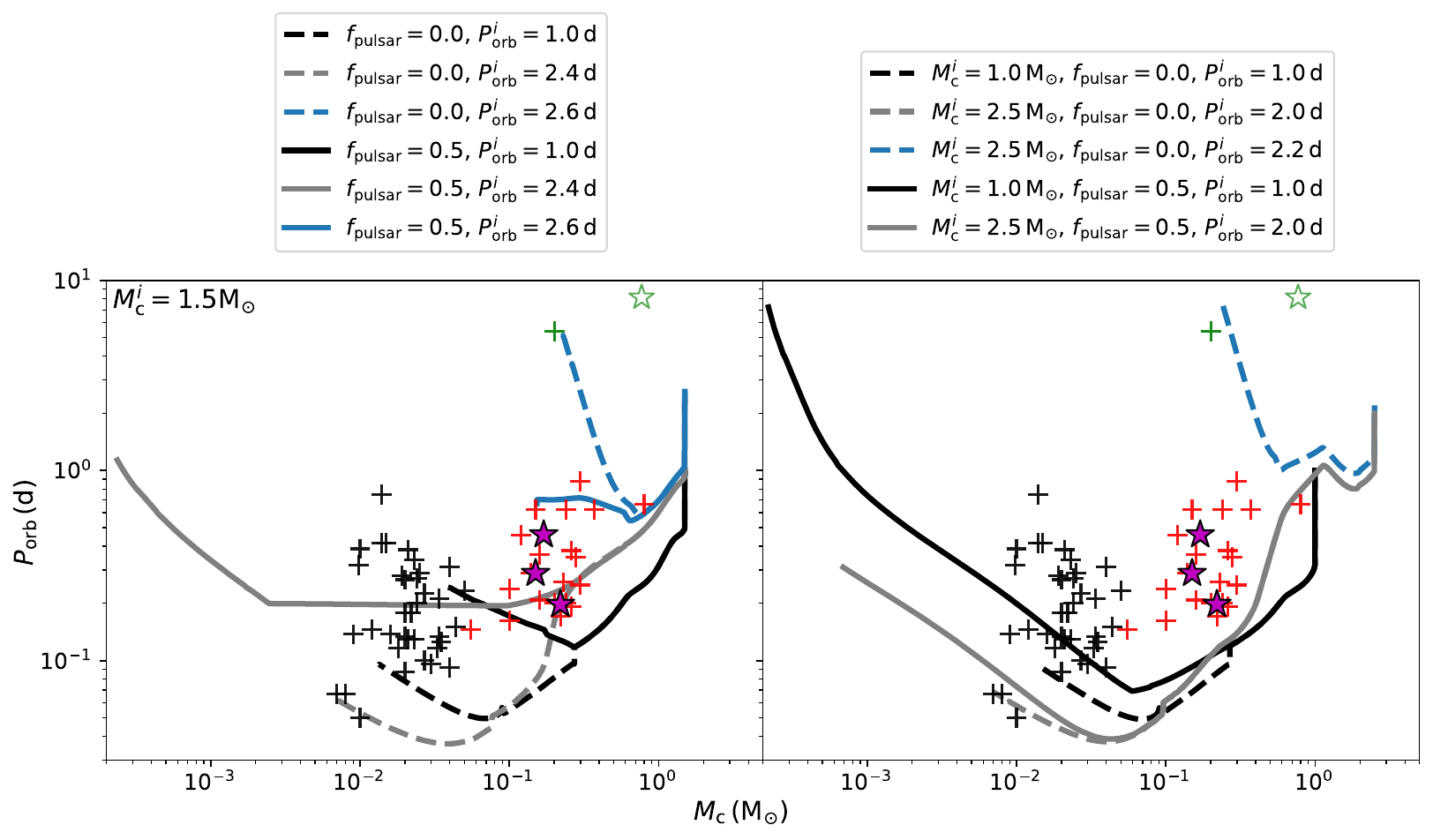}
\caption{Orbital period evolution versus companion mass of selected binaries from our simulated grid shown in Figure\,\ref{fig:beta03}. The term $M_{\rm c}$ is the mass of the companion to the NS, and $P_{\rm orb}$ is the orbital period. The figure shows the effect of pulsar wind irradiation, comparing $f_{\rm pulsar}=0.0$ (dashed lines) and 0.5 (solid lines). The left panel shows the tracks with $M^{i}_{\rm c}=1.5\,\Msun$, $P^{i}_{\rm orb}=1.0$, 2.4, and 2.6\,d. The right panel shows two initial companion masses, $M^{i}_{\rm c}=1.0$ (with $P^{i}_{\rm orb}=1$\,d) and $2.5\,\Msun$ ($P^{i}_{\rm orb}=2.0$ and 2.2\,d). We compared our tracks to $M_{\rm c,min}$ and $P_{\rm orb}$ for BWs (faint black crosses), RBs (faint red crosses), and huntsmen spiders (faint green crosses), from \citet[][and references therein]{Nedreaas_master_thesis}. The green star is the huntsman candidate 2FGL\,J0846.0+2820, observed by Fermi-LAT \citep{2017ApJ...851...31S}. The magenta stars show the three transitional MSPs (tMSPs) observed, \devi{PSR\,J1023+0038} \citep[$M_{\rm c}=0.22\,\Msun$ and $P_{\rm orb}=4.75$\,hr;][]{2019MNRAS.488..198S,2005AJ....130..759T}, IGR\,J18245–2452 \citep[$M_{\rm c,min}=0.17\,\Msun$ and $P_{\rm orb}=11.03$\,hr;][]{2013Natur.501..517P}, and XSS\,J12270-4859 \citep[$M_{\rm c,min}=0.15\,\Msun$ and $P_{\rm orb}=6.91$\,hr;][]{2015ApJ...800L..12R,2015MNRAS.454.2190D,2014MNRAS.441.1825B}.}
\label{fig:bw_eg1}
\end{figure*}

\begin{figure}[!ht]
\centering
\includegraphics[width=\linewidth]{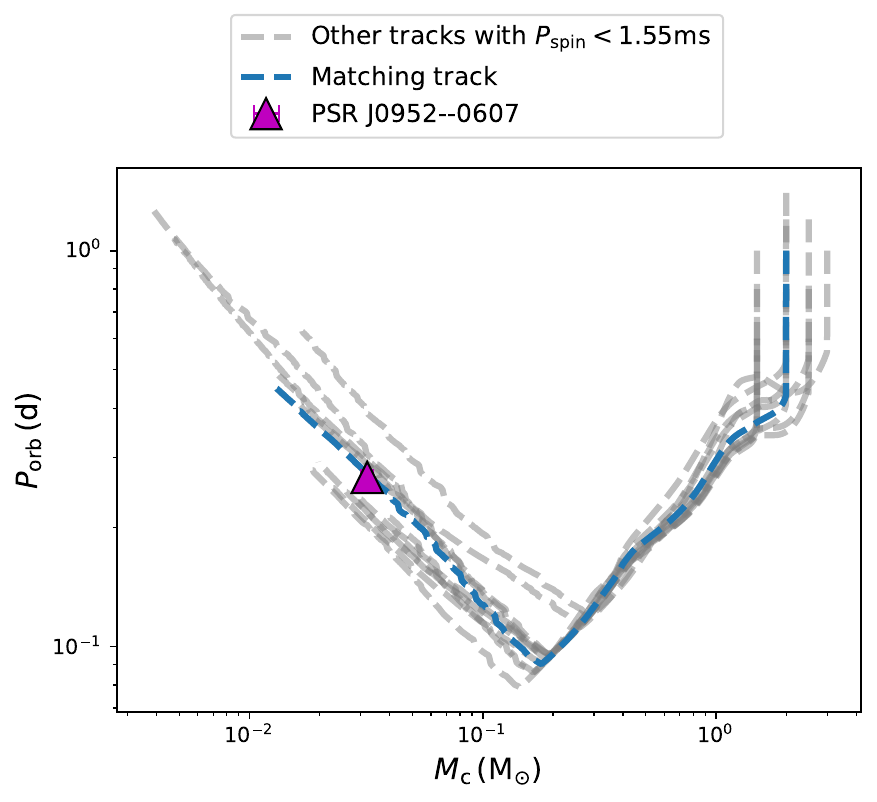}
\caption{Orbital period versus companion mass for diverging binary tracks from our simulated grids that have a similar NS spin as the observed PSR\,J0952--0607 (1.41\,ms) within a 10\,\% margin (dashed gray curves). The term $M_{\rm c}$ is the mass of the companion to the NS, and $P_{\rm orb}$ is the orbital period. All the tracks \devi{shown} have $\beta=0.3$ and $f_{\rm pulsar}=0.5$. We further show the binary that best matches the sources in its orbital evolution in blue curve, which has initial companion mass of 2.0\,\Msun, initial NS mass of 1.3\,\Msun, and initial orbital period of 1.0\,d.}

\label{fig:BW_eg}
\end{figure}

\subsection{Effect of pulsar wind irradiation}
\label{sec:results:irrad_grid}

Figure\,\ref{fig:beta03} shows the evolutionary grids with an $M^{i}_{\rm NS}=1.3\,\Msun$ and varying pulsar wind efficiency ($f_{\rm pulsar}$) \devi{and we see the need for strong irradiation in order to explain all of the observed spiders}. We also show the initial parameter space for spiders in Figure\,\ref{fig:BWParam_space_f1}, their dynamical end, as well as the final orbital periods in the color bar, for $f_{\rm pulsar}=0.0$ and 0.5 and initial NS masses $1.3\,\Msun$. Many of the binaries shown in Figure\,\ref{fig:BWParam_space_f1} (around 50 to 60\% in all simulated spiders) end up suffering a dynamical instability (shown by \devi{triangle} symbols in the figure) \devi{ divided into two categories; those that become unstable during the first RLO phase and those that become unstable toward the end of the second RLO phase. For these binaries in the second category, the NS still manages to accrete enough material to spin it up to milliseconds during the preceding case A RLO. Hence, these could be seen as spider binaries until the onset of instability, after which they would be seen as isolated spinning pulsars. Figure\,\ref{fig:bw_eg2} shows the total mass-loss rates and the mass-transfer rates of the companion versus time of three types of interacting binaries (with pulsar wind irradiation) we have studied so far: case A , near-$P_{\rm bif}$, and case B binaries. The binaries have initial orbital properties as $M^{i}_{\rm c}=1.5\,\Msun$ and  $M^{i}_{\rm NS}=1.3\,\Msun$, and three values for $P^{i}_{\rm orb}=$1.0, 2.4, and 2.6\,d. Going from low irradiation ($f_{\rm pulsar}=0.05$) to high ($f_{\rm pulsar}=0.5$), we see the mass-loss rates spike to very high values for case A and near-$P_{\rm bif}$ binaries, which could instigate an instability. }


\devi{We discussed the occurrence of dynamical instability toward the end of the second RLO phase in the previous section. The presence of irradiation would only increase the occurrence of this instability}. Irradiation removes matter from the surface of the companion star, causing rapid stellar expansion. Increased $f_{\rm pulsar}$ results in a higher number of dynamical unstable systems (goes from 50\% to 62\% in simulated spiders when going from 0.0 to 0.5 in $f_{\rm pulsar}$). We include these dynamically unstable systems in our study since by the time the instability onset happens, the average MSP binary is already $\sim 10^\devi{10}\rm\,yrs$ old and could have been observed during the preceding MSP phase. Hence, there is a reasonable chance to observe these binaries as BW binaries within a Hubble time ($1.4\times 10^{10}\rm\,yrs$). For \devi{this} part of the parameter space the companion star would be destroyed due the instability. Therefore, it could provide an additional channel to the population of Galactic isolated MSPs (20\% of the total Galactic MSP population) that is observed \citep{2005AJ....129.1993M}, with previous RLO spinning up the NS and subsequent merging destroying the companion. \devi{The idea that isolated MSPs could originate from BWs irradiating away their companions has been previously suggested both theoretically \citep{1988Natur.334..227V} and through observations of MSPs with planets/asteroid belts \citep{2009ApJ...703.2017W,2013ApJ...766....5S}.}

Going from $f_{\rm pulsar}=0.0$ to $f_{\rm pulsar}=0.5$ in the converging systems (Figure\,\ref{fig:BWParam_space_f1}), the final orbital periods widen significantly and this effect depends on the initial properties of binary. The widest orbits are seen with $M^{i}_{\rm c}=1.0\,\Msun$, where the final orbital periods are $\gtrsim 10$\,d. For higher companion masses, the final orbital periods lie in the range of 0.1 to 1\,d. Figure\,\ref{fig:bw_eg1} shows the orbital evolution corresponding to three $M^{i}_{\rm c}$ (1.0, 1.5, and 2.5\,\Msun) and $f_{\rm pulsar}=0.0$ to $f_{\rm pulsar}=0.5$, for various values of $P^{i}_{\rm orb}$. For case A RLO binaries ($P^{i}_{\rm orb}=1.0$\,d; shown by black lines), we see increased orbital widening with $f_{\rm pulsar}=0.5$. The companions to the NS become fully convective around 0.2\,\Msun, after which the pulsar wind is turned on. The orbit responds to the mass loss by expanding depending on $f_{\rm pulsar}$. Going from $M^{i}_{\rm c}=1.0\,\Msun$ to $1.5\,\Msun$ (for $f_{\rm pulsar}=0.5$) the expanding orbits cover a larger part of the observed BW space. This is due the fact that all of these binaries are undergoing RLO on a nuclear timescale. For a fixed NS mass and orbital period, more massive companions (or higher $q$) lead to more accretion onto the NS and a higher spin-down luminosity (and correspondingly high orbital expansion). However, with an $M^{i}_{c}=4.0\,\Msun$, the accreted matter decreases down to about $0.004\,\Msun$ as the binary suffers a dynamical instability soon after the first RLO-onset. This is significantly less than $1.0\,\Msun$ accreted with $M^{i}_{c}=3.0\,\Msun$. Hence, there is a limit to how high in companion mass we can go with a fixed initial NS mass and reproduce BWs. We can see this clearly by looking at Figure\,\ref{fig:BWParam_space_f1}. \devi{The effect of irradiation on RLO can be seen in Figure\,\ref{fig:bw_eg2} (top row) with increasing mass loss after the first RLO phase with increasing $f_{\rm pulsar}$.  }

Next we look at the tracks for $P^{i}_{\rm orb}=2.4$\,d in Figure\,\ref{fig:bw_eg1}, corresponding to the gray curves (left panel). For $M^{i}_{\rm c}=1.5\,\Msun$, the effect of $f_{\rm pulsar}=0.5$ is significant on the orbital evolution. The near-$P_{\rm bif}$ binary maintains a relatively wide orbit (around 0.2\,d) crossing the RB regime. \devi{With increasing $f_{\rm pulsar}$, the duration of the mass-transfer phase is shortened as majority of mass loss is due to irradiation (see Figure\,\ref{fig:bw_eg2}). The radio pulsar wind will start as long as the two conditions described in Section\,\ref{sec:methods:eject_regime} are met. Hence, strong irradiation disrupts the RLO phase as the companion detaches more easily from its Roche lobe due to higher mass loss. This does not significantly reduce the number of spiders binaries produced because most of the accretion happens early on in the evolution. In Figure\,\ref{fig:ucxb_eg1}, we see the NSs spin-up to milliseconds toward the start of RLO, for all the binaries presented.} Going to $M^{i}_{\rm c}=2.5\,\Msun$ and $P^{i}_{\rm orb}=2.0$\,d (gray curves in right panel), the near-$P_{\rm bif}$ tracks no longer pass through the RB region with the heavier companion (irrespective of $f_{\rm pulsar}$). For intermediate-mass companions ($\gtrsim 2.0\,\Msun$), the initial angular momentum loss by RLO dominates, \devi{as magnetic braking is not operating}. For very short durations, the mass-transfer rate could even breach the Eddington limit, which results in any excess material being lost from the system taking away even more angular momentum. This continues until the star reaches around $1.0\,\Msun$ (and 1\,d in Figure\,\ref{fig:bw_eg1}) and magnetic braking \devi{turns on}. The following evolution is governed by magnetic braking losses and behaves similar to case A RLO, during its first RLO phase. The fast initial mass loss causes the initially $2.5\,\Msun$ star to become fully convective around $0.2\,\Msun$, after which it follows a UCXB-like evolution. Hence, initial companion masses $\gtrsim2.0\,\Msun$ in near-$P_{\rm bif}$ binaries do not reproduce RBs (also shown in Figure\,\ref{fig:BWParam_space_f1}). If a heavy NS is present ($\sim 2.0\,\Msun$), it could be argued that correspondingly heavier companions ($\gtrsim 2.0\,\Msun$) could form RBs. However, NSs with masses $1.8\,\Msun$ and above very easily cross the maximum NS mass threshold and very few binaries in those grids would be viable spiders (see Section\,\ref{sec:results:acc_eff}).

We can determine also the evolutionary history of PSR\,J0952--0607, the fastest spinning Galactic BW so far (with spin 1.41\,ms), from our simulations, considering $\beta=0.3$ and $f_{\rm pulsar}=0.5$. To account for uncertainties during mass transfer and in our calculation of $P_{\rm spin}$, we consider a 10\% margin and take all binary tracks with $P_{\rm spin}\lesssim1.55$\,ms. The tracks are shown in Figure\,\ref{fig:BW_eg}. We compare the simulated tracks to the estimated orbital parameters, $M_{\rm NS}=2.35^{+0.17}_{-0.17}\,\Msun$, $M_{\rm c}=0.032^{+0.002}_{-0.002}\,\Msun$, and $P_{\rm orb}=6.42$\,hr \citep{2022ApJ...934L..17R}, and select the best matching track (within the errors). The best matching track started with $M^{i}_{\rm NS}=1.3\,\Msun$, $M^{i}_{\rm c}=2.0\,\Msun$ and $P^{i}_{\rm orb}=1.0$\,d (also shown in Figure\,\ref{fig:BWParam_space_f1}). The binary underwent case A RLO and ended due to dynamically unstable RLO from a fully convective companion; the final NS mass was 2.35\,\Msun\, and final NS spin 1.2\,ms. Hence, strong irradiation of the companion explains this source.

\begin{figure}[!ht]
\centering
\includegraphics[width=\linewidth]{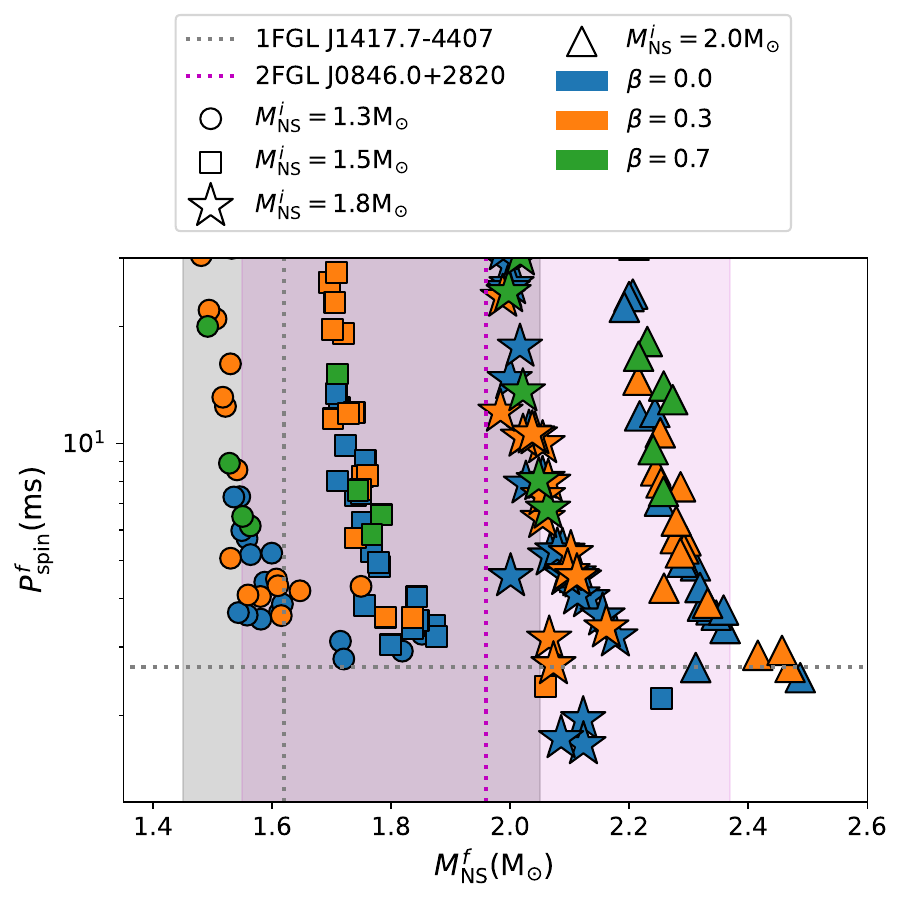}
\caption{ \devi{Final NS spins ($P^{f}_{\rm spin}$) versus final NS masses ($M^{f}_{\rm NS}$) in diverging binaries for $\beta=0.0$ (blue colored symbols), $\beta=0.3$ (orange colored symbols), and $\beta=0.7$ (green colored symbols). We group the binaries by the initial NS mass ($M^{i}_{\rm NS}$) corresponding to different symbols (see legend). We also compare to the observed spin and estimated pulsar mass of the huntsman source 1FGL\,J1417.7--440 \citep[2.66\,ms and $1.62^{+0.43}_{-0.17}\,\Msun$, shown by the gray dashed lines and gray shaded area for the uncertainties;][]{2016ApJ...820....6C,2018ApJ...866...83S} and the estimated pulsar mass of the huntsman candidate 2FGL\,J0846.0+2820 \citep[$1.96^{+0.41}_{-0.41}\,\Msun$, shown by the vertical pink dashed lines and pink shaded area for the uncertainties;][]{2017ApJ...851...31S} .} }
\label{fig:HTspin_hist}
\end{figure}

\begin{figure*}[!ht]
\centering
\includegraphics[width=\linewidth]{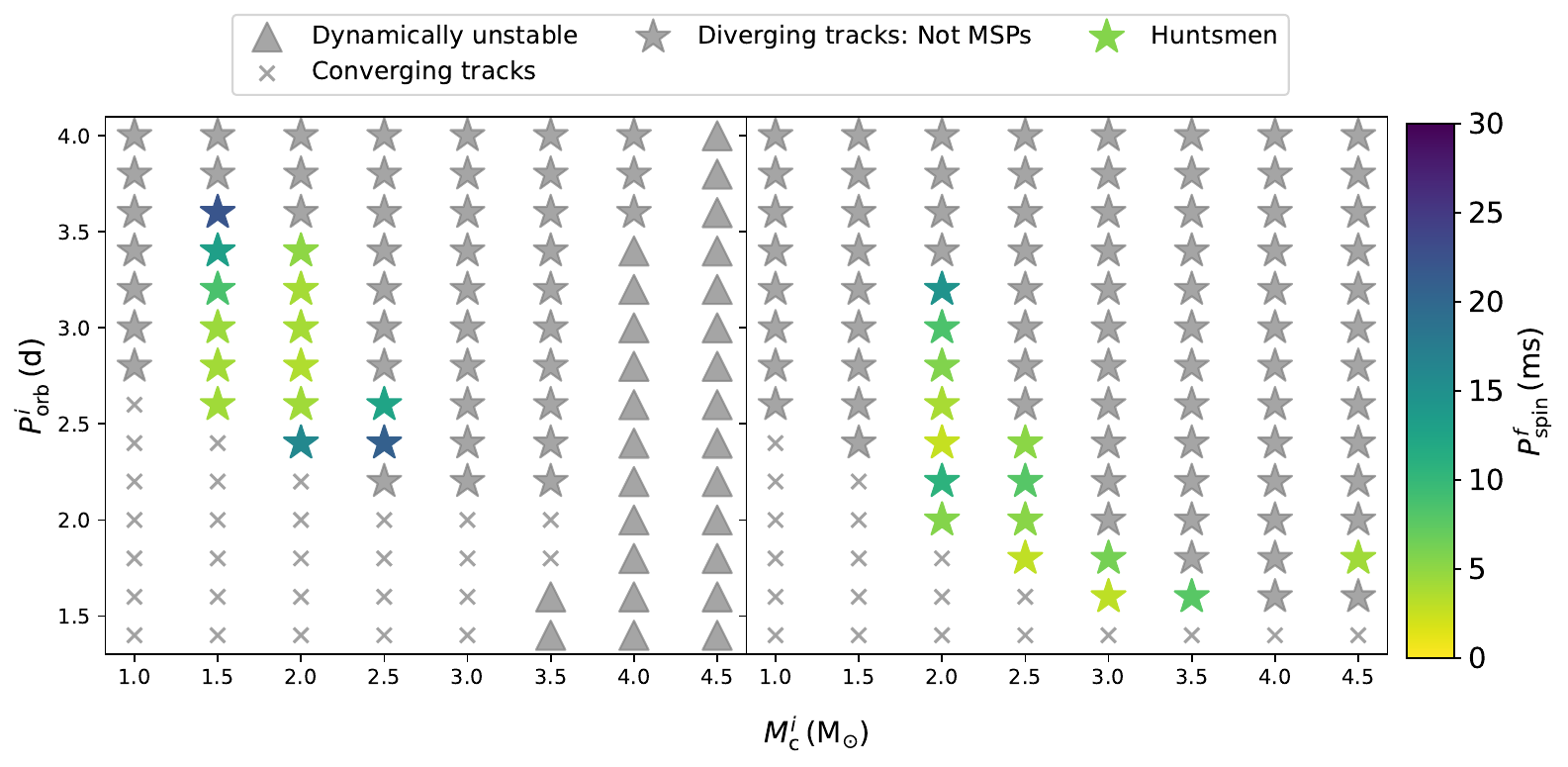}
\caption{Allowed initial parameter space to form huntsmen spiders ($f_{\rm pulsar}=0.0)$ for an initial NS of mass $1.3\,\Msun$ (left) and $2.0\,\Msun$ (right). The term $M^{i}_{\rm c}$ is the initial mass of the companion to the NS, and $P^{i}_{\rm orb}$ is the initial orbital period. The color bar shows the final NS spins ($P^{f}_{\rm spin}$), and the gray stars denote the systems that formed diverging orbits but did not spin up to the MSP range. We show the end states of the tracks, if they encountered a dynamical instability (triangle symbols) or ended with one of the criteria for successful termination (star symbols). Converging tracks are shown by gray crosses.} 
\label{fig:HTParam_space}
\end{figure*}

\begin{table*}[htp]
\begin{center}
\begin{threeparttable}
\caption{Observed and estimated properties of the two observed huntsmen spiders.} 
\begin{tabular}{lllllll}
\hline
Huntsman name & $M_{\rm c,min}$ & $P_{\rm orb}$ & $P_{\rm spin}$ & $M_{\rm NS}$ & $M_{\rm c}$ & $e$ \\ \hline\hline
\multicolumn{1}{c|}{1FGL\,J1417.7--440} & 0.2\,\Msun\tnote{1} & 5.37\,d\tnote{3} & 2.66\,ms\tnote{1} & $1.62^{+0.43}_{-0.17}\,\Msun$\tnote{2} & $0.28^{+0.07}_{-0.03}\,\Msun$\tnote{2} & 0.0 \\
\multicolumn{1}{c|}{2FGL\,J0846.0+2820} & - & 8.13\,d\tnote{4} & - & $1.96^{+0.41}_{-0.41}\,\Msun$\tnote{4} & $0.77^{+0.20}_{-0.20}\,\Msun$\tnote{4} & 0.06\tnote{4} \\
\hline
\end{tabular}
\label{tab:HT_obs}
\tablefoot{The term $M_{\rm c,min}$ is the minimum mass, calculated assuming a pulsar mass of $1.4\,\Msun$ and inclination of $90\degree$; $P_{\mathrm{orb}}$ is the orbital period; $P_{\rm spin}$ is the observed pulsar spin; $M_{\mathrm{NS}}$ is the estimated pulsar mass; $M_{\mathrm{c}}$ is the companion mass observed from Fermi-LAT; and $e$ is the observed eccentricity.}
\begin{tablenotes}
\item[1]\citet{2016ApJ...820....6C},\item[2]\citet{2018ApJ...866...83S},\item[3]\citet{2015ApJ...804L..12S},\item[4]\citet{2017ApJ...851...31S}.
\end{tablenotes}
\end{threeparttable}
\end{center}
\end{table*}

\subsection{Huntsman spiders}
\label{sec:results:HT}

Huntsman spiders are considered to comprise of a red giant and a pulsar in a widening orbit \citep{2015ApJ...804L..12S,2019ApJ...872...42S}. They evolve from a case B RLO phase in LMXBs. In terms of companion masses, they are similar to RBs, with the difference being that their orbits are an order of magnitude wider than RBs. Looking at Figure\,\ref{fig:beta03}, we see that the binary evolution tracks reproduce the huntsmen regime irrespective of the value of $f_{\rm pulsar}$. Contrary to RBs and BWs, with a high $f_{\rm pulsar}$ the orbits do not widen with increasing $f_{\rm pulsar}$. Figure\,\ref{fig:bw_eg1} (left panel, with $M^{i}_{\rm c}=1.5\,\Msun$ and $P^{i}_{\rm orb}=2.6$\,d, shown by blue curves) shows that wider orbits correspond to $f_{\rm pulsar}=0.0$. Since these binaries are undergoing case B RLO, the companions to the NS are giant stars that have developed a deep convective envelope. Convective envelopes expand rapidly on mass loss and drive a high mass-transfer rate. When the radio pulsar wind turns on and irradiates the companion, it further increases the instantaneous mass loss from the companion. The companion detaches from its Roche lobe, pausing the RLO phase. We see the shortened RLO phase in Figure\,\ref{fig:bw_eg2} (bottom row). The resulting effect is a decreased expansion of the orbits.

We also find more conservative mass-accretion rates better reproducing the only current huntsman system with an observed spin period \citep[2.66\,ms for the source 1FGL\,J1417.7--440;][]{2016ApJ...820....6C}. Table\,\ref{tab:HT_obs} describes the various properties of this huntsman source, as well as the other huntsman known so far, 2FGL\,J0846.0+2820. \devi{Figure\,\ref{fig:HTspin_hist} shows the final spin period distributions versus the final NS masses of binary tracks that end up in the huntsmen regime, corresponding to the three $\beta$ values investigated in our study (0.0, 0.3, and 0.7). Compared to BWs and RBs, more conservative mass transfer ($\beta<0.3$) is required for an NS with an initial mass of $1.3\,\Msun$ to spin up to 2.66\,ms.  With a $\beta=0.7$, the minimum spin period reached is 5.83\,ms.} With a $\beta=0.3$, only NSs with initial masses $\gtrsim1.5\,\Msun$ are able to reproduce the pulsar spin of 1FGL\,J1417.7--440. We also compare our simulations to the estimated masses of the pulsars in 1FGL\,J1417.7--440 \citep[$1.62^{+0.43}_{-0.17}$;][]{2018ApJ...866...83S} and 2FGL\,J0846.0+2820 \citep[$1.96^{+0.41}_{-0.41}\,\Msun$;][]{2017ApJ...851...31S}. \devi{If the binaries had fully conservative RLO ($\beta=0.0$), for a huntsman to have an NS with mass greater than 2.0\,\Msun, it should have $M^{i}_{\rm NS}>1.3\,\Msun$ (with 1.3\,\Msun, the maximum mass reached is 1.85\,\Msun). Therefore, a huntsman with a confirmed high mass pulsar would have started RLO with a pulsar heavier than the canonical mass of 1.3\,\Msun. }

\begin{figure}[!ht]
\centering
\includegraphics[width=\linewidth]{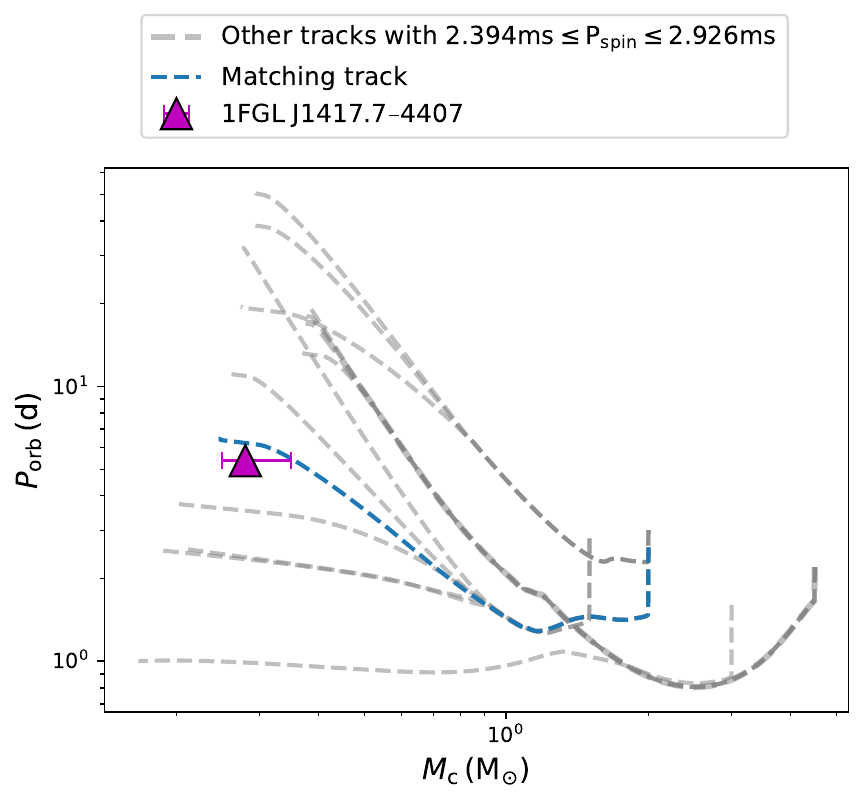}
\caption{Orbital period versus companion mass for diverging binary tracks from our simulated grids that have a similar NS spin as the observed huntsman source 1FGL\,J1417.7--440 (2.66\,ms) within a 20\,\% margin (dashed gray curves). The term $M_{\rm c}$ is the mass of the companion to the NS, and $P_{\rm orb}$ is the orbital period. We further show the binary that best matches the sources in its orbital evolution in blue curves, where $\beta=0.0$ ,$f_{\rm pulsar}=0.1$, $M^{i}_{\rm c}=2.0\,\Msun$, $M^{i}_{\rm NS}=1.3\,\Msun$, and $P^{i}_{\rm orb}=2.6$\,d.}
\label{fig:HT_eg1}
\end{figure}

The observed and estimated properties of the two currently known huntsmen spiders are described in Table\,\ref{tab:HT_obs}. The spin of 1FGL\,J1417.7--440 (2.66\,ms) can be used to constrain its evolutionary history. To reproduce the observed spin and estimated current pulsar mass with an initially canonical NS, we need a fairly conservative RLO phase ($\beta\lesssim 0.3$; see Figure\,\ref{fig:HTspin_hist}). We also use an upper limit of 2.926\,ms and a lower limit of 2.394\,ms on the final spin to better identify the matching binaries, to account for 20\% uncertainty in our calculations for the spin. Though pulsar irradiation is not necessary to reproduce the orbital properties of huntsman spiders, since it does have an effect on the orbit, we compare the different values of $f_{\rm pulsar}$. In Figure\,\ref{fig:HT_eg1} we show the binary tracks that satisfy our selection criteria. We also compare these to the estimated orbital properties of 1FGL\,J1417.7--440 \citep[mass of $0.28^{+0.07}_{-0.03}\rm\,\devi{\Msun}$ and orbital period 5.37\,d;][]{2018ApJ...866...83S,2015ApJ...804L..12S}, and see what tracks pass through the observed properties. We find that the binary track that best matches the source's observed orbital properties (within the errors) corresponds to $\beta=0.0$ and $f_{\rm pulsar}=0.1$, and with the initial parameters, $M^{i}_{\rm c}=2.0\,\Msun$, $M^{i}_{\rm NS}=1.3\,\Msun$, and $P^{i}_{\rm orb}=2.6$\,d. The final spin of the pulsar is 2.73\,ms, which is within 3\% of the observed spin. This binary ends as a He-WD with mass 0.25\,\Msun\, and a $1.73\,\Msun$ NS in an orbit of 6.46\,d.

As for the source 2FGL\,J0846.0+2820, since we do not know its spin, there is a wider range of binary tracks that could be used to explain it, and almost all $f_{\rm pulsar}$ values can reproduce its orbital parameters (see Figure\,\ref{fig:beta03}). Compared to 1FGL\,J1417.7--440 (and other spider systems), 2FGL\,J0846.0+2820 has a non-negligible eccentricity in its orbit \citep[0.06;][]{2017ApJ...851...31S}. This contradicts the general assumption that these systems are old enough for tidal forces to have efficiently circularized the orbits \citep{1995A&A...296..709V}. This observed eccentricity could be the result of the interaction with a circumbinary disk \citep{2014ApJ...797L..24A} or inefficient tides \citep[such as in radiative envelopes, which would be the case for MS stars;][]{2002MNRAS.329..897H}.

Although we can reproduce the orbital parameters observed, we also need to study which diverging LMXB/IMXB systems can spin up NSs to MSP spins. Figure\,\ref{fig:HTParam_space} shows the allowed initial parameter space for the spin-up of NSs in diverging LMXBs/IMXBs. The presented figure shows grids of initial pulsar masses of 1.3 and $2.0\,\Msun$, for $\beta=0.3$ and $f_{\rm pulsar}=0.0$. We present the figures with $\beta=0.3$ instead of 0.0 as we leave room for uncertainties during the RLO phase that were not accounted for in our simulations. The color bar highlights the region of the initial parameter space where the NS was able to spin up to $\leq 30\rm\,ms$. Not all diverging binaries will be able to spin up to MSP range. As we go from a less massive NS to a more massive NS, the peak of the spin distribution (systems with the maximum spin-up due to accretion) moves to higher companion masses and lower orbital periods, from centered around $2.0\,\Msun$ and 2.5--3.5\,d to being centered around $2.5\,\Msun$ and 1.6--3.0\,d. 

The boundary of the huntsmen parameter space depends on the initial $q$ and $P^{i}_{\rm orb}$ of the binary (see Figure\,\ref{fig:HTParam_space}). The range of acceptable $q$ is almost constant for both NS masses and corresponds to regions of high accreted mass. If the companion to the NS is not massive enough, the mass-transfer rate will be too low for efficient NS spin-up. If it is too massive, the RLO phase is very short as the star evolves quick. If the RLO phase is super-Eddington, accretion would be further limited. The top boundary of the parameter space is limited due to the evolutionary state of the companion. Wider orbits than $\sim 3.5$\,d results in the companion filling its Roche lobe after it has developed a deep convective envelope (late case B RLO), making the RLO phase short. In Figure\,\ref{fig:HTParam_space} (right panel, with $M^{i}_{\rm NS}=2.0\,\Msun$), we also see an isolated binary with $M^{i}_{\rm c}=4.5\,\Msun$ and $P^{i}_{\rm orb}=1.8$\,d that has an NS with MSP-like spin. The final binary is a wide He-WD/NS binary (final WD mass $0.31\,\Msun$ and final period 33\,d) with a final NS mass of $2.26\,\Msun$. The neighboring binaries do not drive high enough mass-transfer to spin up their NS, and we did not find a similar binary with a lower initial NS mass.

\subsection{Transitional millisecond pulsars}
There is also a sub-category of compact MSPs that have been observed, binaries that switch from the rotation-powered state (where the spin-down of the pulsar dominates the evolution) to accretion-powered state (mass-transfer from the companion dominates) and vice-versa \citep[see review by][]{2022ASSL..465..157P}. These binaries are called transitional MSPs (tMSPs), and so far there are 3 observed, \devi{PSR\,J1023+0038} \citep{2009Sci...324.1411A}, IGR\,J18245–2452 \citep{2013Natur.501..517P}, and XSS\,J12270-4859 \citep{2014MNRAS.441.1825B}. The companion masses for all of them are in the range 0.15 to 0.2\,\Msun\, and orbital periods 4 to 11\,hr \citep[][and references therein]{2022ASSL..465..157P}, which put them in the RB regime of the observed parameter space. Figure~\ref{fig:bw_eg1} shows the three tMSPs along with some of the tracks from our simulated binaries. Two of these lie the near-$P_{\rm bif}$ tracks for $M^{i}_{\rm c}=1.5\,\Msun$ (gray lines in the figure), where the tracks differentiate for $f_{\rm pulsar}=0.0$ and 0.5. The point of differentiation of the two tracks is the turning on of the pulsar wind, and it explains the transitioning of the MSP binaries from accretion-powered to rotation powered state. Some other factors that should be considered to explain tMSPs are accretion disk instabilities \citep{2001NewAR..45..449L}, which are not yet fully understood. 

\section{Discussion}
\label{sec:discussion}

\subsection{Spider binary formation}

We can compare our simulated spiders to those from previous works. While we find the formation of RBs and BWs to be linked, \citet{2013ApJ...775...27C} found RBs and BWs to come from two distinct populations of LMXBs, depending on the irradiation efficiency. For BWs, they found $f_{\rm pulsar}\lesssim0.08$ to be the required irradiation efficiency, while for RBs, it was $f_{\rm pulsar}\gtrsim0.08$. They explored mainly case A RLO systems, $M^{i}_{\rm c}$ around $1.0\,\Msun$, $M^{i}_{\rm NS}$ around $1.4\,\Msun$, and $P^{i}_{\rm orb}$ from 0.08 to 1.4\,d, and found increased orbital expansion with increasing $f_{\rm pulsar}$. Qualitatively, we also see similar behavior in our results (see Section\,\ref{sec:results:irrad_grid}). Particularly for BWs, looking at similar systems in Figure\,\ref{fig:beta03}, for $f_{\rm pular}=0.0$ to 0.05, we cover the lower-left region of the observed BW space. The degree to which the orbit expands for $f_{\rm pular}\gtrsim 0.1$ is lesser than \citet{2013ApJ...775...27C}. \devi{Using different criteria for the start of the radio pulsar phase than \citet{2013ApJ...775...27C}, we see the phase start at similar evolutionary stages for similar systems (at donor mass of $0.2$--$0.3\,\Msun$ for the case\,A XRBs).}

For RBs, \citet{2013ApJ...775...27C} found them to be produced from case A binaries ($\sim 1$\,d) with strong irradiation ($f_{\rm pulsar}\gtrsim 0.08$). In our work, the majority of RBs are produced at much wider initial orbits (near-$P_{\rm bif}$ at RLO onset, starting at around 2.0 to 2.4\,d at ZAMS), along with high irradiation. With $f_{\rm pulsar}=0.5$, we also see rapid expansion of the case A binaries (solid black curve and $M^{i}_{\rm c}=1.5$ in the right panel of Figure\,\ref{fig:bw_eg1}). However, this expansion in orbits is not enough to bring them completely into the RB regime. \citet{2014ApJ...786L...7B} and \citet{2015ApJ...798...44B} studied the effect of X-rays on the companion surface, as well as pulsar wind irradiation. They found that all BWs should have passed through a RB stage previously. We find that while not all BWs passed through the RB region, most binary tracks passing RB tracks ended up in the BW region of the observed parameter space, and the BWs that passed through the RB region, were H-surface deficient and in narrow orbits. The number of observed BWs to RBs observed is similar (approximately 63\% and 37\%, respectively), which implies their evolutionary lifetimes are comparable. For the BWs and RBs in our simulated tracks, by the time the spider phase begins the binaries are $\gtrsim 10^{9}$\,yr old, as most of the prior evolution happened during the MS phase, which is long for low-mass stars.

The need for a persistent magnetic braking was also suggested by \citet{2020MNRAS.495.3656G} and \citet{2021MNRAS.500.1592G}, where they found that magnetic braking must remain active even when the companion has developed a fully convective structure and the pulsar wind has turned on. While the authors focused their works on BWs, they suggested that the pulsar wind is too weak to remove enough mass from the pulsar companions. They looked at companions with initial masses of $1.0\,\Msun$. Their conclusion is similar to our results, as we also find the binary tracks not reproducing all of the observed BWs for initial companion masses of $1.0\,\Msun$ (see $f_{\rm pulsar}=0.5$ in right panel of Figure\,\ref{fig:bw_eg1}). With increasing companion masses, we cover more of the observed BW parameter space. The MSP source PSR\,J1953+1844 (in the globular cluster M71) that was recently discovered, seems to be the evolutionary link between tidarrens and RBs \citep{2023Natur.620..961P}. It has an orbital period of 53.3\,min and an estimated companion mass of 0.07\,\Msun. Since it is associated with a globular cluster (and would hence have a lower metallicity) and our work seeks to study Galactic spiders, we do not include it in our observed sources \devi{catalog}. Still, \citet{2023Natur.620..961P} suggested that this source was formed from a similar process as UCXBs (with $P^{i}_{\rm orb}$ values near $P_{\rm bif}$), which is similar to our results, albeit at different metallicities. 


\subsection{Accretion efficiency and non-conservative evolution}
As we saw in Section\,\ref{sec:methods:interact_pulsar}, in the literature non-conservative mass-accretion efficiencies ($\beta \gtrsim 0.5$) are \devi{favored} since they can reproduce the LMXB observations. We find that in order to reproduce the observed properties of spider binaries and the observed spins, more conservative efficiencies are better suited ($\beta \lesssim 0.3$). An assumption of non-conservative mass accretion throughout the entire evolution is proxy for any instabilities that might be affecting accretion \citep[for instance, propeller effects or thermal-viscous instabilities]{1975A&A....39..185I,1981ARA&A..19..137P,1996ApJ...464L.139V}. We include the propeller phase in the NS spin evolution (see Section\,\ref{sec:methods:propel_regime}) by removing 10\% of the transferred mass due to the propeller action. Since this phase is highly uncertain even in its effectiveness, we assume a low efficiency of loss of angular momentum during this phase. We also assumed an full efficiency for transfer of angular momentum from the accreted matter during RLO (see Section\,\ref{sec:methods:acc_regime}). \devi{Our results are similar to those by \citet{2024MNRAS.tmp.2289K}, who found that lower $\beta$ values are required to reproduce accreting MSPs. }

As for accretion disk instabilities, we include a model that describes the presence of a accretion disk above a critical mass-transfer rate, below which an X-ray irradiated disk would be unstable (see Section\,\ref{sec:methods:interact_pulsar}). Since we have already included inefficient mass accretion due to these two processes, the mass-accretion efficiency ($\epsilon=1-\beta$) that is set to a fixed value during the entire binary evolution accounts for any additional causes of mass or angular momentum loss. It also accounts for any uncertainties in our models of inefficient accretion, since we use simplistic approximations. \devi{A binary would evolve similarly due to a high value of fixed $\beta$ compared to a low $\beta$ with restricted accretion at some instances. To test this, we ran \mesa\, models without the X-ray irradiated accretion disk assumptions and with $\beta=0.7$ and 0.3, and $f_{\rm pulsar}=0.0$. The initial properties were $M^{i}_{\rm c}=1.5\,\Msun$, $M^{i}_{\rm NS}=1.3\,\Msun$, and $P^{i}_{\rm orb}=1.0$\,d. We compared it to runs with same initial parameters and with the X-ray irradiated disk assumption, and found no significant differences in the evolution.} 

\subsection{Pulsar spin evolution and sub-millisecond pulsars}

There is a lot of uncertainty in our understanding of pulsar spin evolution, including pulsar magnetic field evolution and the pulsar braking index \citep[see review by][]{2024Galax..12....7A}. The spin down for an isolated pulsar follows a power law that dictates the rate of change in the angular frequency ($\dot{\Omega}$) with the angular frequency ($\Omega$), which is $\dot{\Omega}\propto -\Omega^{3}$. The range of values generally explored for the braking index are from 2.5 to 6.0 \citep{2016ApJ...819L..16A, 2018ApJ...863L..40W}, with values closer to 3.0 considered more canonical \citep{1971ApJ...164..529S,1975ApJ...196...51R, 2005yCat.7245....0M,2015PhRvD..91f3007H}. Still there are observations of the braking index in the wide range of $10^{-6}$ to $10^{6}$ \citep{2004MNRAS.353.1311H,2012MNRAS.420..103B,2012ApJ...761..102Z}.

\begin{figure}
\centering
\includegraphics[width=\linewidth]{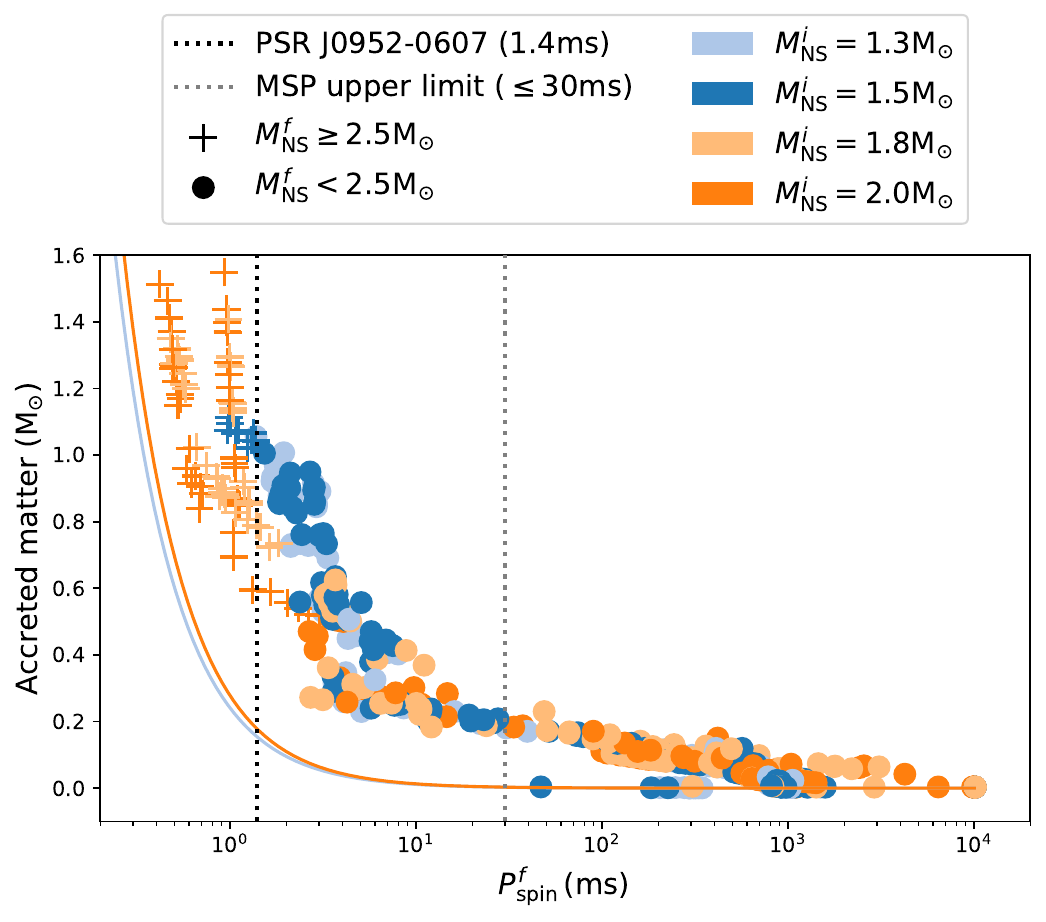}
\caption{\devi{Accreted mass versus final NS spins ($P^{f}_{\rm spin}$) in our simulated grid with $\beta=0.3$. We show the different initial masses as different colors (shown in the legend) and distinguish between the NSs that had $M^{f}_{NS}\ge 2.5\,\Msun$ (crosses) and those where $M^{f}_{NS}< 2.5\,\Msun$ (circles). We also compare our results to the equilibrium spin period calculations from \citet{2012MNRAS.425.1601T}, done for $M^{i}_{NS}=1.3\,\Msun$ (light blue curve) and $2.0\,\Msun$ (dark orange curve). The black dotted line shows the observed spin for PSR\,J0952--0607 (spin period of 1.41\,ms) and the gray dotted line shows the upper limit for MSPs considered (30\,ms).}}
\label{fig:spin_tauris}
\end{figure}

Studies have suggested that MSPs are close to their spin equilibrium, determined by the accretion disk and magnetosphere interaction \citep{2012ApJ...746....9P}. \citet{2012MNRAS.425.1601T} studied the formation of MSPs and derived the equilibrium spin depending on the mass accreted onto the pulsar. They suggested that since most LMXBs undergo non-conservative mass transfer, the MSP can be spun to $\gtrsim 1.4$\,ms. Similarly, in our results, we find a minimum spin of around 1.38\,ms (for $\beta=0.3$; see Section\,\ref{sec:results:acc_eff}), after disregarding systems where the NS crossed the maximum NS mass limit (taken to be $2.5\,\Msun$). Figure\,\ref{fig:spin_tauris} shows the accreted matter onto the NS versus the final NS spin from our simulated grid (for $f_{\rm pulsar}=0.0$). The correlation is compared to the equilibrium spin period calculations from \citet{2012MNRAS.425.1601T}; our simulated NSs require more accreted mass for the same spin-up in the MSP regime. Reasons for discrepancy could be existing uncertainties in spin evolution and accretion physics. We separate out the cases where the final NS mass was greater than 2.5\,\Msun\, (shown as stars in the figure). We find a correlation between the NS spin and the accreted mass, with sub-millisecond spins produced only by NSs beyond the 2.5\,\Msun\, mass limit.

According to Figure\,\ref{fig:spin_tauris}, NSs in LMXBs and IMXBs accrete between 0.2 and $1.0\,\Msun$ to spin up to millisecond periods ($1.38\rm\,ms\leq P^{f}_{\rm spin}\leq 30.0\rm\,ms$). The NSs with initially higher masses ($\gtrsim 1.8\,\Msun$) accrete more than lighter NSs and hence spin up to a higher degree. Naturally, some of these heavy accretors could exceed the NS mass limit and enter the sub-MSP regime. \devi{Accretion induced collapse (AIC) of massive NSs was also studied by \citet{2023ApJ...951...91C} where they found inefficient RLO required initial NS masses of at least 2.16\,\Msun in order for AIC to occure}. \citet{2018A&A...620A..69H} found that masses greater than $2.0\,\Msun$ have an upper limiting frequency of around 810\,Hz (lower $P_{\rm spin}$ limit of 1.23\,ms) depending on the NS equation of state, while lighter NSs could spin faster (up to 1200\,Hz or 0.83\,ms), inconsistent with observations. The spin at which an accreting NS would collapse into a BH also depends on the initial mass: for $M^{i}_{\rm NS}=1.3\,\Msun$, it is at  1.38\,ms and for $M^{i}_{\rm NS}=2.0\,\Msun$, it is at  2.66\,ms. For an NS lighter than $M^{i}_{\rm NS}=1.3\,\Msun$, the switch from NS to BH could occur at a spin lower than 1.38\,ms. While our resulting spins are not as small as those by \citet{2018A&A...620A..69H}, we see a similar trend. For masses $\gtrsim 2.0\,\Msun$ the equation-of-state could change, leading to a collapse of the NS into a BH, or the formation of a hybrid star with sub-MSP spin \citep{PhysRev.172.1325,2000A&A...353L...9G,2017A&A...600A..39B}. \citet{2024ApJ...962...80L} accounted for selection bias due to pulse broadening and suggested chances for Square Kilometer Array to observe sub-millisecond pulsar spins would be negligible. Since there are no current observations of any such sub-MSP star, sub-MSPs may not exist if those NSs have collapsed into BH.

\subsection{Uncertainties in redback formation}

We find an evolutionary connection between UCXBs, RBs, and tiderrans (see Section\,\ref{sec:results}). Our results for the formation of UCXBs are similar to other studies where they find UCXBs originating from LMXBs starting from close to $P_{\rm bif}$ \citep{2002ApJ...565.1107P,2005A&A...440..973V,2005A&A...431..647V,2014A&A...571A..45I,2019RAA....19..110H}. The idea that the Tidarren PSR\,J1311--3430 went through a UCXB phase was also hinted at by \citet{2013ApJ...775...27C} and. There is the fine-tuning problem in the formation of UCXBs. With the standard magnetic braking prescription \citep{1972ApJ...171..565S,1981A&A...100L...7V,1983ApJ...275..713R} often used (also used in our study), there is a narrow range of orbital periods (that is near-$P_{\rm bif}$) and companion masses from which UCXBs can be produced \citep{2005A&A...440..973V,2005A&A...431..647V,2014A&A...571A..45I}. Even though strong irradiation widens near-$P_{\rm bif}$ orbits and some of them might no longer be UCXBs, some UCXBs can still be reproduced (see $f_{\rm pulsar}=0.5$ in Figure\,\ref{fig:beta03}). 

The dependence of $P_{\rm bif}$ on magnetic braking assumptions and the effects on LMXB evolution have been extensively studied \citep{2002ApJ...565.1107P,2004MNRAS.355.1383L,2009ApJ...691.1611M,2016ApJ...830..131C,2021ApJ...909..174D,2022MNRAS.517.4916E,2024MNRAS.tmp.1119E}, and it has been suggested that a stronger magnetic braking process is needed to form UCXBs (and consequently RBs) from a larger initial parameter space. \devi{\citet{2021MNRAS.503.3540C} tested the magnetic braking prescription suggested by \citet{2019ApJ...886L..31V} in the formation of binary MSPs and UCXBs. They found that while efficient magnetic braking does increase the allowed initial parameter space to form UCXBs, wide-orbit binary MSPs are not reproduced.} In regards to spiders, \citet{2020MNRAS.495.3656G} and \citet{2021MNRAS.500.1592G} focused on reproducing the observed BW properties and proposed a sustained magnetic braking prescription that will drive RLO; magnetic braking will not stop operating at $M_{\rm c}\sim 0.2\,\Msun$ and the companion has lost its radiative core. \citet{2023ApJ...950...27G} compared \mesa{} simulations with single star and LMXB observations and also found saturated magnetic braking to best reproduce both types of sources. Still the standard magnetic braking prescription can reproduce LMXBs with periods less than one hour to around ten days, better than some other proposed prescriptions \citep{2024MNRAS.tmp.1119E,2023ApJ...950...27G}. Carrying out a population synthesis study of UCXBs would also tell us what fraction of LMXBs in a stellar population start RLO close to their $P_{\rm bif}$. Hence, to disentangle the various formation channels and types of magnetic braking prescriptions, a combined investigation of detailed simulations and population synthesis analysis would be required, which would further help in understanding the formation of spiders.

\section{Conclusions}
\label{sec:conclusions}

Compact MSP binaries (or spider binaries) are good laboratories for studying binary evolution. In this work, we calculated the NS spin evolution and identified binaries that formed MSPs (using the criteria for NS spins: $P^{f}_{\rm pular}\le 30$\,ms). Assuming a model for an X-ray irradiated accretion disk and three different values of the accretion efficiency (0.0, 0.3, and 0.7), we ran simulation grids spanning four values of initial NS mass ($1.3\,\Msun$, $1.5\,\Msun$, $1.8\,\Msun$, and $2.0\,\Msun$), a range of initial companion masses ($1.0\,\Msun$ to $4.5\,\Msun$), and initial orbital periods (0.6\,d to 4.0\,d). Following the RLO phase, as the pulsar spins down, it loses rotational energy in the form of spin-down luminosity, ablating the outer envelope of the companion star and causing changes in the orbital evolution. We used four different values for the efficiency of this process (0.0, 0.05, 0.1, and 0.5) and presented the allowed initial parameter space to produce BWs, RBs, and huntsmen spiders. After comparing our simulated grids to the observed spider populations, we arrived at the following conclusions:
\begin{itemize}
    \item With the assumption of an unstable accretion disk due to X-ray irradiation during the mass-transfer phase, an accretion efficiency of \devi{at least} 70\% is needed in order to reproduce the observed spins and pulsar masses of BW systems. High accretion efficiency ($>70\%$) also allows for the formation of the huntsmen spiders from a binary with a canonical NS \devi{in a widening orbit during RLO}.
    \item  For BW formation, a high pulsar wind efficiency ($f_{\rm pulsar}=0.5$) \devi{is} required to reproduce the full observed parameter space, although lower efficiencies ($0.0<f_{\rm pulsar}<0.5$) can still reproduce part of the systems. However, the tidarrens (J1311--3430, J1653--0159, and J0636+5128; BWs with orbital periods $\lesssim1.0$\,d and $0.01$M$_{\odot}$ companions) require no irradiation. We also predict that J0636+5128 (similar to its \devi{neighbors}) should not have any hydrogen on its companion's surface, as the companion loses its hydrogen envelope through stable mass transfer. The binary tracks passing through the tidarren region also pass through the RB region, connecting BWs and RBs as part of the same initial population. These binaries start RLO at near-$P_{\rm bif}$ and pass through an UCXB phase. Their evolution is dependent on the magnetic braking prescription used.
    \item  With high pulsar wind irradiation ($f_{\rm pulsar}=0.5$), the orbits of the near-$P_{\rm bif}$ binaries expand drastically, covering most of the RB region. This is true for the companion masses $\lesssim 1.5\,\Msun$. The near-$P_{\rm bif}$ binaries are the main formation channel for RBs. Hence, RBs are subject to the fine-tuning problem seen in UCXBs, where only a narrow range of values for the initial \devi{orbital} parameter space, \devi{with standard magnetic braking}, can reproduce the observations.
    \item We identified the initial parameter space for diverging NS binaries to spin up as MSPs. These systems would form the small population of huntsmen spiders that have been observed. With fully conservative accretion (assuming an X-ray irradiated accretion disk), an NS in diverging orbits with an initial mass of $1.3\,\Msun$ gains a maximum of $0.55\,\Msun$. If a huntsman pulsar is discovered to have a pulsar mass $\gtrsim2.0\,\Msun$, it was most likely more massive than $1.3\,\Msun$ when formed. 
    \item We identified the prior evolution of the pulsars BW PSR\,J0952--0607 and huntsman 1FGL\,J1417.7--440 by matching their observed spins and orbital properties to binary tracks. For PSR\,J0952--0607, high irradiation ($f_{\rm pulsar}=0.5$) after case A RLO is required to reproduce its estimated orbital properties. For 1FGL\,J1417.7--440, the decreased expansion of the orbit due to pulsar wind irradiation ($f_{\rm pulsar}=0.1$) matches the observations well, though irradiation is not required to reproduce the orbital properties of these diverging NS binaries.
    \item The NSs in LMXBs and IMXBs accrete between 0.2 and $1.0\,\Msun$ to spin up to MSP spins. For NSs to attain sub-MSP spins, the mass accreted is large enough to collapse it into a BH ($\gtrsim 1.0\,\Msun$ for $M^{i}_{\rm NS}=1.3\,\Msun$). Additionally, the spin at which an accreting NS would enter the sub-MSP regime (and collapse) increases with increasing mass, going from 1.38\,ms to 2.66\,ms when $M^{i}_{\rm NS}$ goes from $1.3\,\Msun$ to $2.0\,\Msun$. This could explain why sub-MSPs are not observed and why the fastest spinning BW spins at 1.4\,ms. With $\beta=0.3$, we can reproduce the fastest spinning BW currently observed PSR\,J0952--0607 (observed spin of 1.41\,ms and pulsar mass of $2.35\,\Msun$). Hence, fast spinning NSs are key in identifying the boundary between NSs and BHs.
    \item Around 50 to 60\% of our simulated spiders ended up becoming dynamically unstable due to mass transfer initiated from a fully convective companion. These binaries would go through a common envelope phase, and the surviving binaries would be isolated spinning NSs. These binaries could explain the observations of isolated MSPs.
\end{itemize}


Since this study involved running detailed \mesa\, simulations with a fixed set of initial parameters, we did not know the formation probability of these types of systems and how it would compare to the observed populations. In order to calculate the formation rates of spiders, we would need to carry out a population synthesis study. Using large-scale parameter studies, we can better constrain certain physical parameters governing binary evolution. These include processes affecting the evolution prior to the XRB phase (e.g., the supernova mechanism) and more directly involved physical processes (e.g., magnetic braking and pulsar wind irradiation). Since LMXBs have very unstable accretion disks that cause outbursts that would affect pulsar spin-up \citep[for instance, due to X-ray irradiation during RLO;][]{1999MNRAS.303..139D}, unsteady or transient accretion must be further investigated in detail. While we do include a model to account for these effects that would cause non-conservative mass transfer, these instabilities and their effects on the observed populations are highly uncertain and require further investigation. Finally, this study was conducted at solar metallicity because of the substantial population of Galactic spiders that have been observed. However, other metallicities should also be investigated to account for the spiders that have been observed in globular clusters with metallicities below solar \citep[e.g., PSR\,J1953+1844 in the globular cluster M71 at 1/10th solar metallicity; ][]{2023Natur.620..961P}.

\begin{acknowledgements}
The authors thank the anonymous referee for their constructive comments that helped improve the manuscript. The authors thank Thomas Tauris, Hai-Liang Chen\devi{, and Rapha{\"e}l Mignon-Risse}  for their useful discussions. This project has received funding from the European Research Council (ERC) under the European Union’s Horizon 2020 research and innovation programme (grant agreement No. 101002352, PI: M. Linares). C.S.Y. acknowledges support from the Natural Sciences and Engineering Research Council of Canada (NSERC) DIS-2022-568580.
\end{acknowledgements}

\bibliographystyle{aa}
\bibliography{main.bib}

\begin{appendix}

\section{Numerical stability of \mesa\,tracks}
\label{sec:appendix:num_stab}

\begin{figure}[!ht]
\centering
\includegraphics[width=\linewidth]{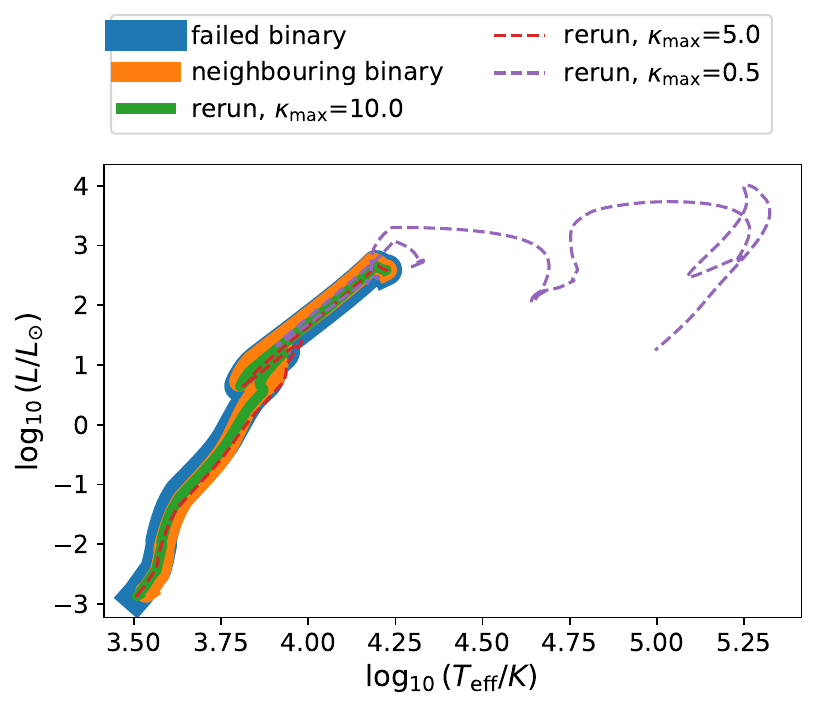}
\caption{Stellar luminosity ($L$) versus effective temperature ($T_{\rm eff}$) of a binary from the simulated grids with an initial companion mass of 4.5\,\Msun, initial NS mass of 2.0\,\Msun, and initial orbital period of 1.4\,d for different maximum opacity values ($\kappa_{\rm max}$). The solid blue curve is the failed binary. The solid green, dashed red, and dashed purple curves are with $\kappa_{\rm max}$ value of 10.0, 5.0, and 0.5, respectively. We compare the tracks to a neighboring track (without any numerical instability) in solid orange.}
\label{fig:HRD_failed_opmax}
\end{figure}

\citet[][Section 5.2]{2022arXiv220205892F} list a wide range of criteria for successful termination of the \mesa\, track, (1) the age of the companion to the NS exceeds the Hubble age (13.8\,Gyr), (2) the companion to the NS becomes a WD, and (3) the binary enters a common envelope. The binaries that did not satisfy the criteria for successful termination (mentioned in Section\,\ref{sec:methods:ini_prop}), were rerun with a limitation on the maximum radiative opacity ($\kappa_{\rm max}$) of the companion. The initial failure rate was 16.5\% in the entire initial simulated grid. Since the stellar opacity influences the internal stellar structure, it prevents the star properties from fluctuating too drastically for \mesa\, to handle. The rerun $\kappa_{\rm max}$ value used by \citet{2022arXiv220205892F} is 0.5\,$\rm cm{^2}\,g^{-1}$, which brought down our failure rate to 0.27\%. However, we find this value to be too restrictive for some stars in our initial parameter space, resulting in a significant deviation in the resulting tracks from the successful tracks in close vicinity. This can be seen in Figure\,\ref{fig:HRD_failed_opmax}, where we show a binary (with $\beta=0.0$) having an initial companion mass of 4.5\,\Msun, an initial NS mass of 2.0\,\Msun, and an initial orbital period of 1.2\,d, which failed due to numerical instability. We also show a \devi{neighboring} binary that avoided numerical instabilities with 4.5\,\Msun, initial NS mass of 2.0\,\Msun, and initial orbital period of 1.4\,d. Decreasing the value of $\kappa_{\rm max}$ to 0.5 results in the stellar properties deviating significantly from previous reruns, from the very beginning of the evolution. The binary that previously should have undergone a dynamical instability with a strict $\kappa_{max}$ does not have a high mass-transfer rate and ends as a WD. Hence, to keep the tracks as similar to the original run as possible (while avoiding numerical instabilities) as well as similar in \devi{behavior} to their close \devi{neighbors}, we reran the failed binaries in two batches. First we used a $\kappa_{\rm max}$ of 10\,$\rm cm{^2}\,g^{-1}$, and then we ran the binaries that failed again with a $\kappa_{\rm max}$ of 5\,$\rm cm{^2}\,g^{-1}$. Our final failure rate was 9.19\%, which is a lot higher than the case of a strict $\kappa_{\rm max}$; however, we compromised by checking the final properties of the binaries, and if they seemed to have an MSP-like spin for the NS ($\le 30$\,ms), we considered the binary successful. 

\section{Low accretion efficiency grid}
\begin{figure*}[!ht]
\centering
\includegraphics[width=\linewidth]{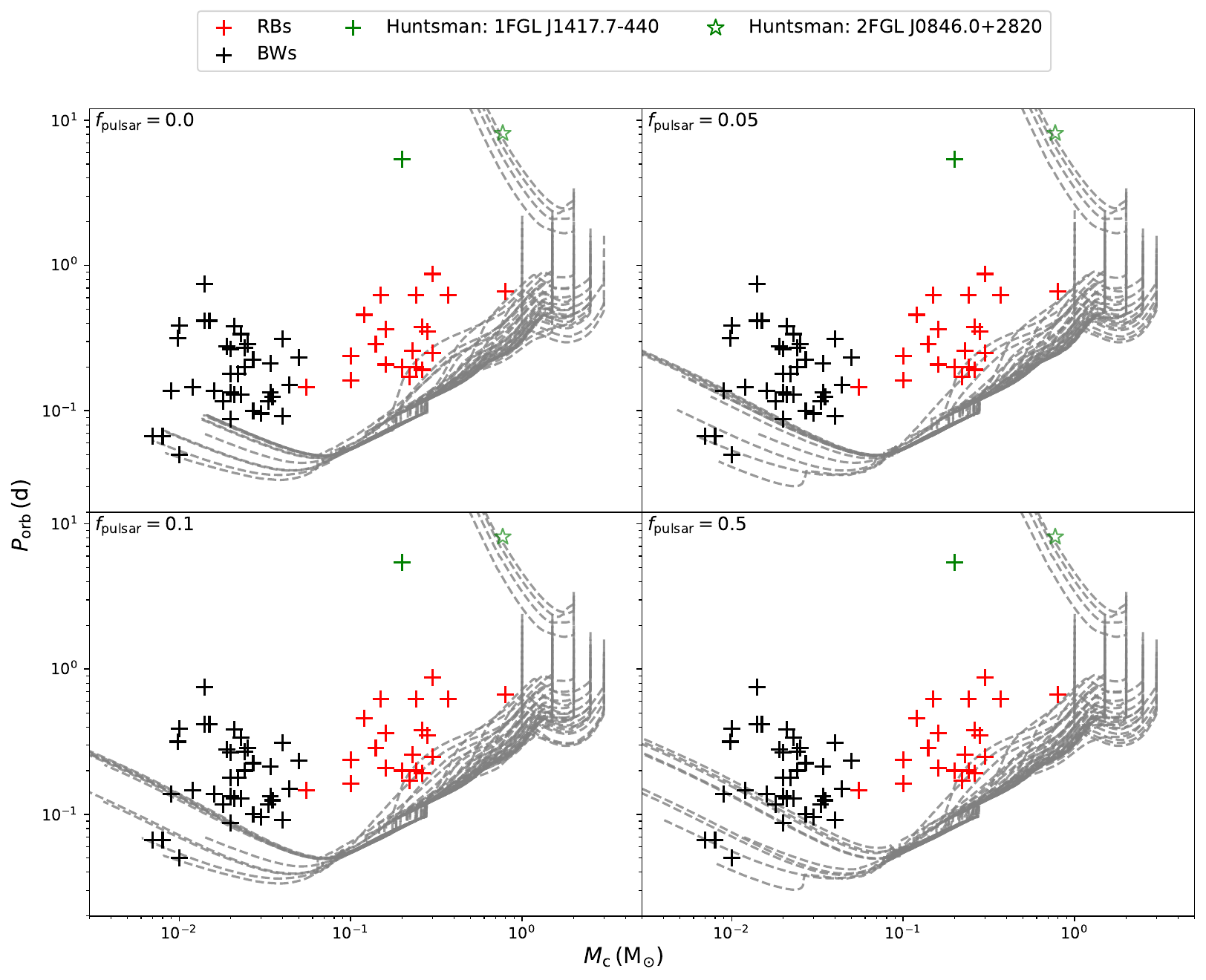}
\caption{Orbital period evolution versus companion mass of the simulated binaries for $\beta=0.7$ and with an initial NS of mass $1.3\,\Msun$ and $f_{\rm pulsar}$ values of 0.0 (top left), 0.05 (top right), 0.1 (bottom left), and 0.5 (bottom right). The term $M_{\rm c}$ is the mass of the companion to the NS, and $P_{\rm orb}$ is the orbital period. We also compare our tracks to estimated $M_{\rm c,min}$ and observed $P_{\rm orb}$ for spiders (Section\,\ref{sec:obs}). The initial binaries have a MS star (range of 1.0 to 4.5\,\Msun) and initial periods of 0.6 to 4.0\,d. }
\label{fig:beta07}
\end{figure*}
We present the evolutionary grids with $\beta=0.7$ in Figure\,\ref{fig:beta07}. The figure also shows the observed population of spiders, BWs, RBs, and huntsmen. As can be seen from the figures, most of the tracks miss the spider populations, and hence we look to more conservative mass-accretion efficiencies for our study.

\end{appendix}

\end{document}